\documentclass[11pt]{iopart}
\usepackage{graphicx}
\usepackage[usenames]{color}

\newcommand{\green}{\color{Green}}

\def \eg{{\em e.g.}}
\def \gtw{\>\hbox{\lower.25em\hbox{$\buildrel >\over\sim$}}\>}
\def \ltw{\>\hbox{\lower.25em\hbox{$\buildrel <\over\sim$}}\>}
\def \cf{{\em cf.}}

\def \cm3{cm$^{-3}$}
\def \radm2{ rad/m$^2$}
\def \dyncm2{dyn cm$^{-2}$}
\def \mach{{\cal M}}

\def \g{\gamma}
\def \be{\begin{equation}}
\def \ee{\end{equation}}
\def \ApJ{{\em Astrophys. J. }}
\def \AJ{{\em Astron. J. }}
\def \MNRAS{{\em Mon. Not. R. Astron. Soc. }}
\def \AandA{{\em Astron. Astrophys. }}

\begin{document}

\title[The dynamic age of Cen A]{The Dynamic Age  of Centaurus A}

\author{Jean A. Eilek$^{1,2}$}
\address{$^1$ New Mexico Institute of Mining and Technology, 
Socorro, New Mexico, 87801, USA}
\address{$^2$ Adjunct astronomer,
National Radio Astronomy Observatory, P.\ O.\ Box O,
Socorro, NM 87801 USA.}
\ead{jeilek@aoc.nrao.edu}



\begin{abstract}

  In this paper I present dynamic models of the radio source Centaurus A,
and critique possible models of in
  situ particle reacceleration (ISR) within the radio lobes.  The
 radio and $\g$-ray data require neither  homogeneous
  plasma nor quasi-equipartition between plasma  and magnetic
  field; inhomogeneous models containing both high-field and low-field
  regions are equally likely. Cen A cannot be as young as the
radiative lifetimes of its relativistic electrons, which range
from a few to several tens of Myr.  Two classes of dynamic models -- flow driven
and magnetically driven -- are consistent with current observations;
  each requires
Cen A to be on the order of a Gyr old.  Thus, ongoing ISR must be occuring 
  within the radio source.  Alfven-wave ISR is probably occuring
throughout the source, and may be  responsible for maintaining the
$\g$-ray-loud electrons.  It is likely to be supplemented by shock
or reconnection ISR which maintains the radio-loud electrons in high-field
regions. 

\end{abstract}

\submitto{\NJP} 
\section{Introduction}

The full story of the radio galaxy Centaurus A is still not known,
despite much careful study over the years.  Cen A is of course worth
study for its own sake, but it also provides a nearby, well-documented
example from which we can understand the dynamic evolution 
of other, similar radio galaxies. In this paper I focus on two
unresolved questions related to Cen A's large-scale radio lobes.

The first question is the age of Cen A.  To answer this we need to understand
the physical state of its radio lobes.
  So-called Fanaroff-Riley Type II (``FRII'') radio galaxies 
 are well understood: 
a collimated, supersonic jet develops a terminal shock where it impacts the
ambient medium.  The shock is apparent as an outer hot spot, and jet plasma
which has been through the shock is left behind as a cocoon.  Ram pressure
drives the growth of the source, as the end of the jet advances into
the ambient medium. However, Cen A
is not an FRII, but a Fanaroff-Riley Type I  (``FRI'';
implicitly defined these days
as any radio galaxy that is not an FRII).  No robust models  exist yet
for the dynamics and development of FRI sources.  The inner jets in some FRI's
have been modelled as turbulent, entraining flows, but that model does not
address the variety of FRI morphologies, nor does it describe
the history and growth of such a source.   More work is needed
before we can understand the dynamics of FRI sources.

The second question is how Cen A keeps shining long past the radiative
lifetimes of its radio-loud and $\g$-ray loud electrons. 
 In this paper I discuss three alternative dynamical models
which may describe Cen A.  All of these models require the 
source to be quite old, with an age between 
several hundred Myr to more than a Gyr. This age range is much 
longer than the electrons' radiative lifetimes, which range from a few
to a few tens of Myr. It follows that the radiative lifetimes
are not a good indicator of the age of the source;  in situ reacceleration
(ISR) must be ongoing within the large radio lobes.  Thus, Cen A
provides a nearby case study of ISR in radio galaxies.

To address these questions,
 I begin in Section  \ref{BasicStuff} by describing the
setting and collecting important facts to be used in the analysis.
In  Section 
\ref{InternalConditions} I review observational constraints on plasma
and magnetic field within the outer lobes.  I argue 
 that inhomogeneous models, containing both strong and weak fields,
are physically likely and consistent with the data.  In Section 
\ref{ModelsThatDontWork} I show that the short radiative
lifetimes require ISR;  models fail which try to grow the source so quickly.
 In Section  \ref{ModelsThatWork} I present three alternative dynamic
models -- buoyant, magnetic tower and flow-driven -- 
 each of which requires the source age to be on the order of a Gyr. 
The buoyant model does not fare well when compared to the data
but both the flow-driven and magnetically-driven models seem consistent with
current observations.  In Section  \ref{InSituReacceleration} 
I critique competing ISR  models. I speculate that Alfven-wave ISR
is operating throughout the lobes and probably maintains the $\g$-ray-loud
electrons, while the radio-loud electrons are energized by shock or
reconnection ISR in high-field regions within the source.  
Finally in Section  \ref{Conclusions} I summarize my arguments
and discuss what this exercise reveals about the astrophysics of Cen A.
I relegate necessary but tedious details to the Appendices.

\begin{figure*}[htb]
{\center
\includegraphics[width=0.95\columnwidth]{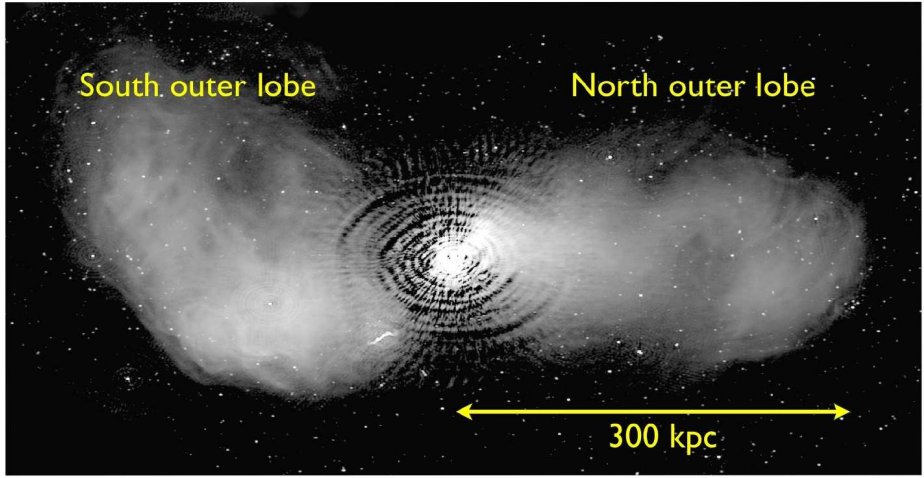}}
\caption{The large-scale structure of Centaurus A, displayed with north
to the right.  Neither lobe shows any sign of limb brightening; the volume
of each lobe appears to be filled with localized, radio-bright filaments
and loops.  Elliptical rings in the center of this image are not physical, 
but are part of the interferometer response (to the
very strong central source) which
has not been fully removed from the image.   The 
inner lobes  sit within and are obscured by these
artifacts.  Mosaiced 1.4-GHz image using ATCA and Parkes data, $60'' \times
40''$ resolution (data are publicly available on the NASA Extragalactic
Database;  see also Feain \etal 2011).}
\label{Fig:OuterLobes}
\end{figure*}

\section{Useful facts and numbers}
\label{BasicStuff}

Cen A's parent galaxy, NGC 5128, is the dominant galaxy in a 
small group of galaxies.   It hosts a massive black hole which is
currently active, and must have  been so for most of the past Gyr. 
The plasma 
driven out by the active galactic nucleus (AGN) has created the 
large radio galaxy which we see today.

\subsection{Large scales:  the outer radio lobes}
\label{LargeScales}

Cen A is perhaps best known by its large-scale, double-lobed radio
source, extending $\sim 8.2^{\circ}$ on the sky, or 540 kpc end-to-end
in projection (using 3.8 Mpc distance;  Harris \etal 2010), as shown
in Figure \ref{Fig:OuterLobes}.  These outer lobes have been called giant
lobes in the literature, but they are not unusually large compared to
other double-lobed radio galaxies.  I therefore drop the "giant" and
just call them "outer lobes" or "radio lobes".

The outer lobes have been detected in radio from 5 MHz to 5 GHz
(\eg, Alvarez \etal 2000), and recent WMAP images detect the outer 
lobes up to 60 GHz (Hardcastle 2009, ``H09'').
While generally similar, the radio structures of the north and
south outer lobes differ in detail.  The north lobe is straighter
(extending $\sim 4.5^{\circ} \sim 300$ kpc in projection), while the south
lobe broadens and bends. The outer lobes are also 
clearly detected as extended $\g$-ray sources (Abdo \etal 2010, 
``A10''), interpreted as inverse Compton scattering (ICS) of photons from
the cosmic microwave background (CMB).

When numbers are needed I illustrate with the north lobe. The angle
$\theta$ by which the north lobe deviates from the sky plane is
unknown; I keep $\mu = \cos \theta$ as a parameter. 
The width of the north lobe is also uncertain.  Although relatively
narrow at high radio frequencies (\eg, Junkes \etal 1993), recent
data suggest 
that it is broader when seen in low radio frequencies (PAPER data at
148 MHz from Stefan \etal 2013;  MWA data at 118 MHz from 
McKinley \etal 2013) and even broader when seen 
at $\g$-rays (recent {\em Fermi}-LAT data from
Yang \etal 2012, ``Y12'').   Based on these data  I treat the north
lobe as a cylinder, with  length $\sim 300 / \mu$ kpc,
and diameter $\sim 180$ kpc.
The geometry of the south lobe is less clear; it may be a
broader ovoid, or it may be a longer structure seen in projection.

\subsection{Small scales:  drivers and inner radio structure}
\label{InnerDrivers}

In the past few Myr  the AGN in NGC 5128 has created
a bipolar inner radio lobe structure, $\sim 15$ kpc end to end (in
projection).   Estimates of the power currently being supplied by
the AGN to each side of the radio source range from 
$\sim 0.7 \times 10^{43}$ erg/s to $\sim 2 \times 10^{43}$ erg/s
(Croston \etal 2009, Wykes \etal 2013, Neff \etal 2014).

Although Cen A is typical of the general radio galaxy population on 
large and small scales -- both showing bipolar radio lobes -- it is atypical
on the transition scales ($\sim 20\!-\!50$ kpc).  No
radio jets have been detected connecting the inner and outer radio lobes.
A previous suggestion of a weak jet extending to NE (Morganti \etal 1999) 
is not supported by newer data (Neff \etal 2014).

\subsection{Conditions in the ambient medium}
\label{IGM_conditions}

The large angular size of Cen A makes direct X-ray measurements of the
surrounding intergalactic medium (IGM) challenging. On small scales 
-- probably representative of the inner part of the IGM surrounding
Cen A -- Kraft \etal (2009) modelled X-rays from a  $\sim 35$-kpc region
surrounding NGC5128 as coming from a sphere with density $\sim
10^{-3}$cm$^{-3}$ and temperature $\sim 0.35$ keV.  

To characterize the IGM on larger scales, we must look for analogs
among  other galaxy groups and poor clusters.  Cen A sits in a small group of
galaxies (Karachentsev \etal 2007) with gravitating mass $\ltw 9
\times 10^{12} M_{\odot}$.  The group is comparable to, but a bit
smaller than, other galaxy groups and poor clusters for which good
X-ray data is available.  The temperature of the IGM has been
measured, typically at $\sim 1$ keV, for quite a few similar galaxy
groups (e.g., Finoguenov \etal 2006, Sun \etal 2009).  When
temperature structure can be resolved, the IGM tends to be cooler in
the core, rising to $\gtw 1$ keV by $\sim 100$ kpc.  Determining the
density of the IGM in faint, irregular galaxy groups is harder.
  Doe \etal (1995) studied several poor clusters, a few
times more massive than the Cen A cluster but of comparable size.
They found the IGM density $\sim 5 \times 10^{-4} \!-\! 3 \times
10^{-3}$\cm3 inside $\sim 100$ kpc, falling off 
$\propto r^{-1.5}$ at larger radii. Thus it seems likely that the
IGM  around most of the outer lobes in Cen A
is less dense than the estimate from Kraft 
\etal (2009) for the inner region of NGC5128, perhaps $n_{IGM} \sim
10^{-4}$\cm3  (see also discussion in O'Sullivan \etal 2013).

This estimate for the  IGM density differs from that of Bouchard
\etal (2007), who searched for HI in 18 galaxies in the Cen A group.
They  detected HI in some but not all galaxies, and ascribed
the non-detections to stripping 
by the IGM, suggesting $n_{IGM} \sim 10^{-3}$\cm3.  However, 
following the basic stripping argument of Gunn and Gott (1972)
with the galaxy properties assumed by Bouchard \etal, it is easy to show
that a  typical galaxy will be stripped by any IGM density
$n_{IGM} \gtw 10^{-4}$\cm3.  

Putting these results together,
  I will scale the  IGM density around the outer lobes of Cen A
to $ 10^{-4}$\cm3,   the temperature to 1 keV, 
and estimate the IGM  pressure as
$ p_{IGM} \sim 3.2 (n_{-4} T_{keV}) \times 10^{-13}$ dyn-cm$^{-2}$.
The likely density decay in the IGM between 100 and 300
kpc  means  $p_{IGM}$ is probably a bit higher 
around the inner part of the radio lobes,
 and a bit lower around the outer part.

\section{Conditions within the outer lobes}
\label{InternalConditions}

In order to evaluate models  of the outer lobes, we must understand
their internal state.   Our diagnostics
are radio and $\g$-ray observations.  
Gamma rays detected  by {\em Fermi}-LAT from the outer lobes of Cen A
are a direct probe of highly relativistic electrons in the outer lobes.
Electron energies $\g \sim 3.4 \times 10^5 \!-\!2.8 \times 10^6$ are needed 
to make $\g$-rays at $100$ MeV $\!-\! 7$ GeV by ICS on the CMB.

Radio emission detected from Cen A  is a separate,  less
direct, probe of the plasma and magnetic field in the lobes. 
Synchrotron emission maps the
electron energy to a photon energy, $\nu_{sy} = a \g^2 B$, where
$a \simeq 0.3 ( 3 / 4 \pi) ( e / m^3 c^5) \sin \theta $ (Pacholczyk 1970; 
noting the maximum of the $F(x)$ synchrotron kernel is at $x \sim 0.3$). 
Numerically, with $\sin \theta \sim 0.5$, this becomes $\nu_{sy} \sim 0.6 
\g^2 B_{\mu G}$.  Thus, radio emission at $10$ MHz comes from electrons
with $\g \sim 4.1 \times 10^3 / B_{\mu G}^{1/2}$; emission at $ 5$ 
GHz comes from electrons with $\g \sim  9.1 \times 10^4 / B_{\mu G}^{1/2}$.
This is a key point: for $\mu$G or stronger magnetic fields, 
{\it the radio emission comes from
a different electron population than does the $\g$-ray emission.}

In this section I use the radio and $\g$-ray data to derive  constraints
on the plasma and field in the radio lobes.  I discuss various
B field models, based both in observations and theory, and 
argue that the magnetic field must vary -- on both small and large scales
-- within the source.  I 
 summarize the possible field values and models
in Table \ref{OL_Bfields}.

\subsection{Relativistic electrons in the radio lobes}
\label{TotalElectronEnergies}

From the observed $\g$-ray and radio power we can derive the total
energy in $\g$-ray-loud electrons and find a useful estimate of the
total energy in radio-loud electrons.

\paragraph{Gamma-ray loud plasma.}

The ICS energy loss rate for an electron at energy $\g$, sitting in the CMB,
 depends only on its energy and fundamental parameters, 
\be
P_{ICS}(\g) \simeq { 4 \over 3} c \sigma_T \g^2 {B_{CMB}^2 \over 8 \pi} 
= b_{ICS} \g^2
\label{Single_particle_power_ICS}
\ee
where $B_{CMB} = (8 \pi u_{CMB})^{1/2} \simeq 3.24 \mu$G.
To find the total ICS luminosity, $L_{ICS}$,
let the $\g$-ray-loud electrons have a power-law energy
distribution, $n_{ICS}(\g) \propto \g^{-t}$.   An electron
spectral index $t \sim 4.2$ will create a $\g$-ray photon index
$ \sim 2.6$ (consistent with the values reported in A10, Y12).
 These ``ICS electrons'' have  total energy
\be
U_{e,ICS} = \int_{\g_2} \g m c^2 n_{ICS}(\g) d \g
\label{ICS_electrons}
\ee
for $\g_2 \sim 3.4 \times 10^5$.  It is straightforward to connect $U_{e,ICS}$
to the ICS luminosity above 100 MeV, 
$L_{ICS} = \int_{\g_2} b_{ICS} \g^2 n(\g) d\g$:
\be
U_{e,ICS} \simeq { t -3 \over t -2} { m c^2 \over b_{ICS} \g_2}
L_{ICS} 
\label{GammaRayLoudElecEnergy}
\ee
To apply this to Cen A, I convert the  $\gamma$-ray  flux from A10 to
luminosity $L_{ICS}$, using $t = 4.2$.  The 
results are given in the first two columns of Table \ref{OL_energies}.

\begin{table}[htb]
\caption{\label{OL_energies}
The energy in relativistic electrons in the outer lobes of 
Cen A.  ICS derivation assumes  electron spectrum
$n_{ICS}(\g) \propto \g^{-4.2}$ in $\g$-ray-loud 
energy range.  Synchrotron derivation assumes 
$n_{sy}(\g) \propto \g^{-2.4}$ in radio-loud energy range
and  observing  frequency $\nu_o = 1.4$ GHz.  
$U_{e,sy}(\nu_1)$ assumes $\nu_1 = 10$ MHz;
$U_{e,sy}(\g_1)$ assumes $\g_1 = 10$.  For other values,
$U_{e,sy} \propto \nu_1^{-0.2} \propto \g_1^{-0.4}$ .
}
\begin{indented}
\item[]
\begin{tabular}{l c c  c c c c }
\br 
\rule{0pt}{16pt} \hspace{-8pt} 
Region   &$L_{ICS}$ &  $U_{e,ICS}$ &   $S(\nu_o)$ 
& $U_{e,sy}(\nu_1) $ & $U_{e,sy}(\g_1)$
\\
     & (erg s$^{-1}$)  &    (erg) &  (Jy) & (erg) & (erg) 
\\[4pt]
\mr
\rule{0pt}{16pt} \hspace{-8pt} 
North lobe  & $5.6 \times 10^{40}$  & $6.6 \times 10^{54}$ 
 &166 & $ 1.4 \times 10^{57} B_{\mu G}^{-1.5} $ 
        & $ 1.6 \times 10^{58} B_{\mu G}^{-1.7} $
\\[4pt]
South lobe  & $8.0 \times 10^{40}$ & $1.0 \times 10^{55}$ 
 &316 & $ 2.7 \times 10^{57} B_{\mu G}^{-1.5} $
      & $ 3.0 \times 10^{58} B_{\mu G}^{-1.7} $
\\[4pt]
\br
\end{tabular}
\end{indented}
\end{table}


\paragraph{Radio-loud plasma.}

Unlike ICS, the synchrotron energy loss rate for an electron at $\g$ 
depends on the local B field: 
\be 
P_{sy}(\g) \simeq { 4 \over 3} c
\sigma_T \g^2 { B^2 \over 8 \pi}  = b_{sy} \g^2 B^2 
\label{Single_particle_power_synch}
\ee 
Let the radio-loud electrons also have a power-law energy distribution,
$n_{sy}(\g) \propto \g^{-s}$  over some energy range $\g_1 \ltw \g 
\ltw \g_{br}$, with an index $s \sim 2.4$ (corresponding to
radio spectral index  $\alpha_{rad}
 \sim 0.7$, typical of the outer lobe spectrum
below a few GHz;  Alvarez \etal 2000, also H09).
  The  total energy in ``synchrotron electrons'' is
\be
U_{e,sy} = \int_{\g_1}^{\g_{br}} \g m c^2 n_{sy}(\g) d \g
\label{synch_electrons}
\ee

Just as for $\g$-ray-loud electrons,
the energy in radio-loud electrons can be connected to the radio power.
However, unlike the ICS result in \eref{GammaRayLoudElecEnergy}, 
the conversion for radio is  subject
to uncertainties in the B field and the low-energy cutoff $\g_1$.
It is convenient here write the synchrotron luminosity in
 terms of the spectrum $S(\nu)$, as $L_{sy} = \int_{\nu_1}^{\nu_{br}} S(\nu)
d \nu$.    A bit of algebra connects the energy in radio-loud electrons
to the B field:
\be 
U_{e,sy} =
{ 2 \over s-2} { m c^2 a^{1/2} \over b_{sy} }
{ \nu_o^{(s-1)/2}  \over    \nu_1^{(s-2)/2}} 
{S(\nu_o) \over B^{3/2}}
\label{RadioLoudElecEnergy}
\ee
Here, $\nu_o$ is the fiducial (observing) frequency, $\nu_1 = a \g_1^2 B$,
and the B field is 
assumed to be approximately constant within the radio-loud region. 

Equation \eref{RadioLoudElecEnergy} contains a
critical parameter, $\nu_1 \propto \g_1^2$.  The literature
contains two different treatments of $\nu_1$. (1) Several authors (\eg,
Myers and Spangler 1985, also A10) choose a low $\g_1$ based on ISR
models.  Typical choices are $\g_1 \sim 10 \!-\! 100$ (corresponding to
$\nu_1 \sim 60$ Hz to 6 kHz in a $\mu$G B field, well below the limit of
terrestrial radio observation).  (2) Other authors (\eg, Burbidge 1956,
Pacholczyk 1970) choose $\nu_1$
as the lowest observed radio frequency, which implies a significantly
 larger $\g_1$.   Because $U_{e,sy}$ is
dominated by the lowest electron energies, this more conservative
choice gives lower values for pressure and energy.  

To calculate $U_{e,sy}$, I use radio fluxes at 1.4
GHz measured by H09 over the same spatial regions in Cen A
used by A10 to measure the $\g$-ray fluxes.  I found $U_{e,sy}$
for $\nu_1 = 10$ MHz and also for $\g_1 = 10$ (in the spirit of A10,
Y12;  corresponding to
$\nu_1 = 60 B_{\mu G}$ Hz); for other values, $U_{e,sy} \propto \nu_1^{-0.2} 
\propto \g_1^{-0.4}$.  The results are given in Table 
\ref{OL_energies}, where two trends warrant comment.
   The lower electron energy cutoff ($\g_1 = 10$)
gives a factor $\gtw 10$ increase in $U_{e,sy}$;  this difference is due
to the assumption that the power law electron spectrum continues to energies
a factor $\gtw 100$ lower than those observed. 
In addition, comparing $U_{e,ICS}$ to either $U_{e,sy}$ value shows 
 that, for $\mu$G-level B fields, the
bulk of the electron energy is in radio-loud electrons; the energy in
ICS-loud electrons is insignificant.

\subsection{From observations: two useful limits}
\label{UsefulLimits}

The radio luminosity can be used to find  lower limits on the
pressure and on the B field in radio-loud regions.

\paragraph{Minimum  magnetic field.}
We begin with the smallest B  field compatible with the radio
power.  Since the plasma is a synchrotron source, there must be a
finite magnetic field, but it may be weak, $p_B \ll p_{rel}$.
In this limit, if the lobes are in pressure balance with the IGM,
we know $p_{rel} \ltw p_{IGM}$.  To use this limit, we must
connect $U_e \simeq U_{e,sy}$ (from Table
\ref{OL_energies}), to $p_{rel}$.  This requires 
two correction factors,  neither of which
can be predicted {\em a priori};  they must 
be constrained from observations or ancillary data, as follows. 

  (1) The radio-loud
plasma may not fully fill the source.  I use the observed  volume
corresponding to the measured $S(\nu_o)$ and $L_{ICS}$ values
($V_{obs} \sim 1.1 \times 10^{71}$cm$^3$ for north lobe, $ \sim 0.87
\times 10^{71}$cm$^3$ for south lobe; H09, A10), but also allow for a
volume filling factor $\phi$.  Slices taken across
the radio image in Figure \ref{Fig:OuterLobes} show that the outer
lobes of Cen A are not limb brightened, but approximately center
filled.  This, plus the fact that few ``holes'' are seen in the image,
suggests that $0.1 \ltw \phi \ltw 1$ describes Cen A.  
(2)
Relativistic baryons may coexist with the radio-loud leptons.  This is
measured by the ``$k$'' factor: $p_{rel} = (1+k)p_e$.  In a pure lepton
plasma, $k = 0$; in galactic cosmic rays, $k \simeq 100$.  Because the ISR
mechanisms (to be discussed in Section \ref{InSituReacceleration}) will
energize baryons as well as leptons, and because baryons do not suffer
radiative losses, we might expect $k \gg 1$ in Cen A.  
I combine both $k$ and $\phi$  in a pressure scaling factor, $\xi =
(1+k)/\phi$, and carry $\xi$ through the calculations.  My
estimated ranges of $k$ and $\phi$ suggest that
$1 \ltw \xi \ltw 10^3$ may hold for the outer lobes of  Cen A.

With these definitions, the pressure in relativistic particles is
$p_{rel} =  \xi  U_{e,sy} / 3  V_{obs} $
Using $U_{e,sy}$  values from Table \ref{OL_energies}, $p_{IGM}$ from
Section \ref{IGM_conditions},  and arguing $ p_{rel} \ltw p_{IGM}$, 
we find the minimum magnetic field allowed in each of the  outer lobes:
\be
B_{min}^{sy}({\rm North}) \sim 0.06 ~\xi^{2/3}  \mu G ~;
\quad 
B_{min}^{sy}({\rm South}) \sim 0.10 ~ \xi^{2/3}  \mu G
\ee
Here, I take $\nu_1 = 10$ MHz and $ n_4 T_{keV} = 1$ for the IGM;
for other values, $B_{min}^{sy} \propto  ( \nu_1^{0.2} p_{IGM})^{-2/3}
\propto  ( \g_1^{0.4} p_{IGM})^{-2/3}$.
The reader should note that  $B_{min}^{sy}$ 
is not to be confused with $B_{min~p}$, the  field
associated with the minimum-pressure condition, which comes next.

\paragraph{Minimum pressure.}
We can also find the lowest pressure (in relativistic particles
and B field) compatible with the observed synchrotron power.  
This standard calculation, described in many
sources (\eg, Pacholczyk 1970), is also subject to
the uncertainty in the low-energy cutoff of the electron spectrum described 
in Section \ref{TotalElectronEnergies}.\footnote{I follow Pacholczyk (1970)
 in assuming that  $\nu_1$ is known; other
authors, \eg, Myers \& Spangler (1985) do the calculation assuming $\g_1$ is
known. Because $\nu_1$ and $\g_1$ are related by the unknown B field, the
algebra is different for each case;  see Neff \etal (2014) for details.}
  Because a higher $\g_1$ gives a lower
minimum pressure, I assume $\nu_1 = 10$ MHz, equivalent to
$\g_1 \simeq 4000 B_{\mu G}^{-1/2}$. For other values $B_{min~p} \propto
\nu_1^{-0.057}$ with my choice $\alpha = 0.7$. 
The key input from the data is the volume
emissivity of the radio source, which  I took from 
 the Parkes 4.75-GHz map of Cen A (Junkes \etal 1993), kindly provided by 
N. Junkes. (Calculations using measurements of the 1.4-GHz
image in Figure \ref{Fig:OuterLobes} gave similar results).
 I measured the average surface brightness within four regions
 which were roughly equivalent to regions 1, 2, 4,
5, used by H09 and A10, and assumed an average line of sight depth $100'
\sim 110 $kpc.  The results, keeping $\xi$ as an unknown parameter, are
given in Table \ref{OL_derived}.  Two trends deserve comment.  
 We see that $B_{min~p} > B_{min}^{sy}$.  This is
expected because $p_{rel} \sim p_B$ in the minimum-pressure model,
while $p_{rel} \gg p_B$ in the minimum-field model. In addition, if
$\xi \gtw 50$, we have $p_{min~p} \gtw p_{IGM}$.  This suggests
that radio emission may come from localized, high-pressure regions within
the lobes;  I return to this in Section \ref{AlternativeModels}.

\begin{table}[htb]
\caption{\label{OL_derived}
Minimum-pressure values for outer lobes.
Calcuations assumed  $\alpha = 0.7$ and 
$\nu_1 = 10$ MHz; for other values $B_{min~p} \propto \nu_1^{-0.057}$.
The scaling factor $\xi = (1+k)/\phi$; $k$ is the ratio of relativistic
baryons to leptons, and $\phi$ is the filling factor (Section \ref{UsefulLimits}).
}
\begin{indented}
\item[]
\begin{tabular}{c c c c  c  c c }
\br 
\rule{0pt}{16pt} \hspace{-12pt} 
Region & $F_{5~GHz}$ & Box area  & $p_{min~p}$ 
& $ B_{min~p}$    
\\
    &(Jy)   &(amin$^2$)  &   (dyn cm$^{-2}$)  & ($\mu$G)
\\[4pt]
\mr 
\rule{0pt}{16pt} \hspace{-8pt} 
North outer &  30. &$9.2 \times 10^3$    & $2.8 ~\xi^{4/7} \times 10^{-14}$ 
& $0.55 ~ \xi^{2/7}$
\\[4pt]
North inner & 45. &$8.0 \times 10^3$  & $3.8 ~\xi^{4/7} \times 10^{-14}$ 
& $0.64 ~\xi^{2/7} $
\\[4pt]
South inner  & 134. &$2.4 \times 10^4$   &$5.1 ~\xi^{4/7} \times 10^{-14}$ 
& $0.74 ~ \xi^{2/7} $
\\[4pt]
South outer  & 84.  &$1.4 \times 10^4$  &$2.9 ~\xi^{4/7} \times 10^{-14}$ 
& $0.56 ~\xi^{2/7} $
\\[4pt]
\br 
\end{tabular}
\end{indented}
\end{table}

\subsection{From observations:  homogeneous models}
\label{CombineSyICS}

Gamma rays produced by ICS are a potentially powerful tool for
determining the magnetic field, because ICS luminosity is independent
of the B field.  Combining $\gamma$ rays with radio synchrotron can in
principle break the $e\!-\!B$ degeneracy and determine B uniquely,
{\it if synchrotron and ICS come from same electron
  population}. However, as we saw above, in Cen A the ICS and synchrotron
emission come from different electron energies; thus further
assumptions are necessary.

One approach (A10, Y12) assumes a homogeneous source, with a uniform B
field and a single electron population.  In this case the electron
energy distribution, $n(\g)$, must be fine-tuned to produce both the
radio spectrum (power law at low frequencies, with steepening above a
few GHz) and the steeper $\g$-ray spectrum. Specifically, the electron
spectrum must have either two breaks in energy, or one break with a
high-energy curvature.  Given an assumed $n(\g)$  which satisfies
these constraints, it is straightforward to estimate the ratio
$U_{e,sy}/U_{e,ICS}$, and from that estimate the B field.

\begin{figure}[htb]
{\center
\includegraphics[width=.425\columnwidth]{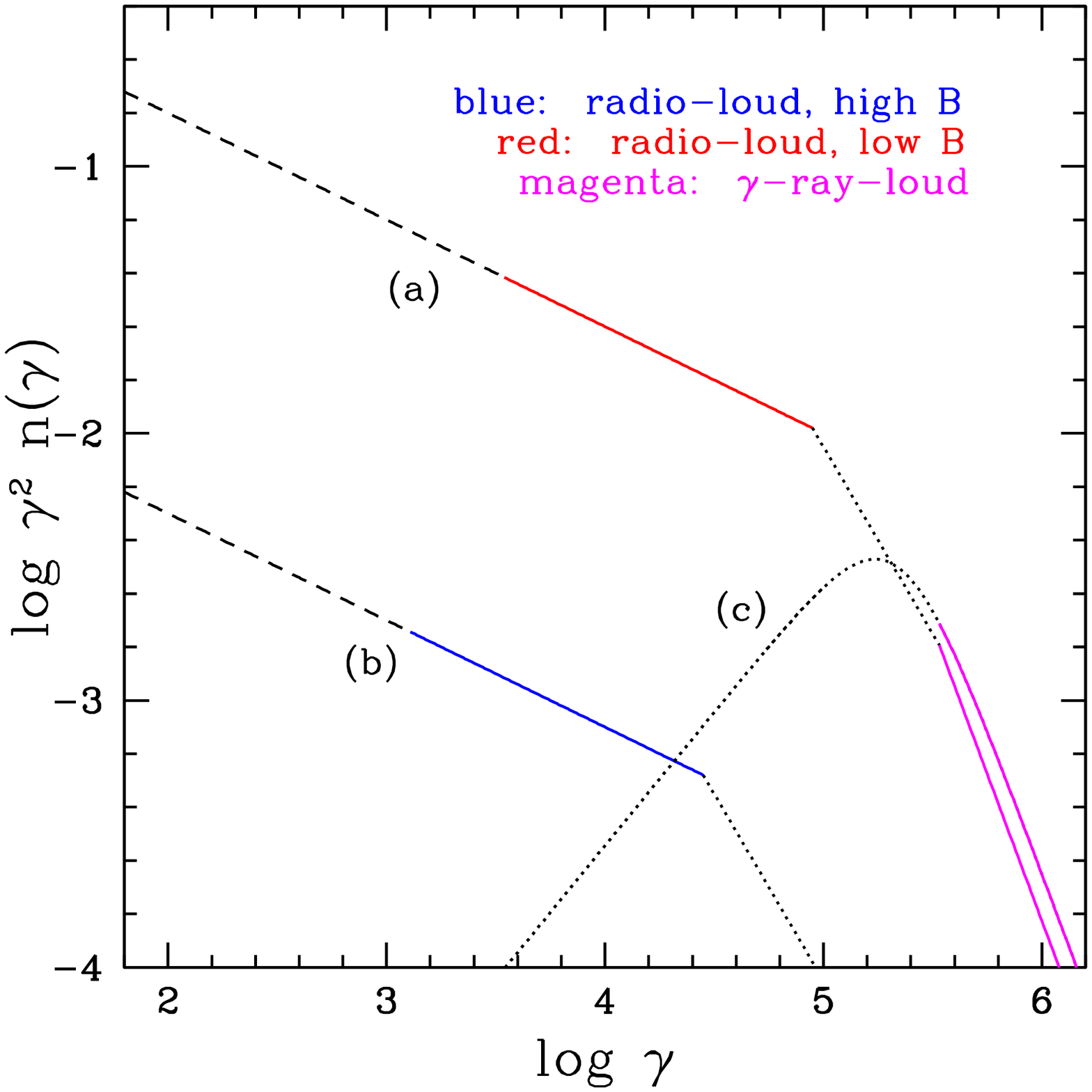}
\includegraphics[width=.425\columnwidth]{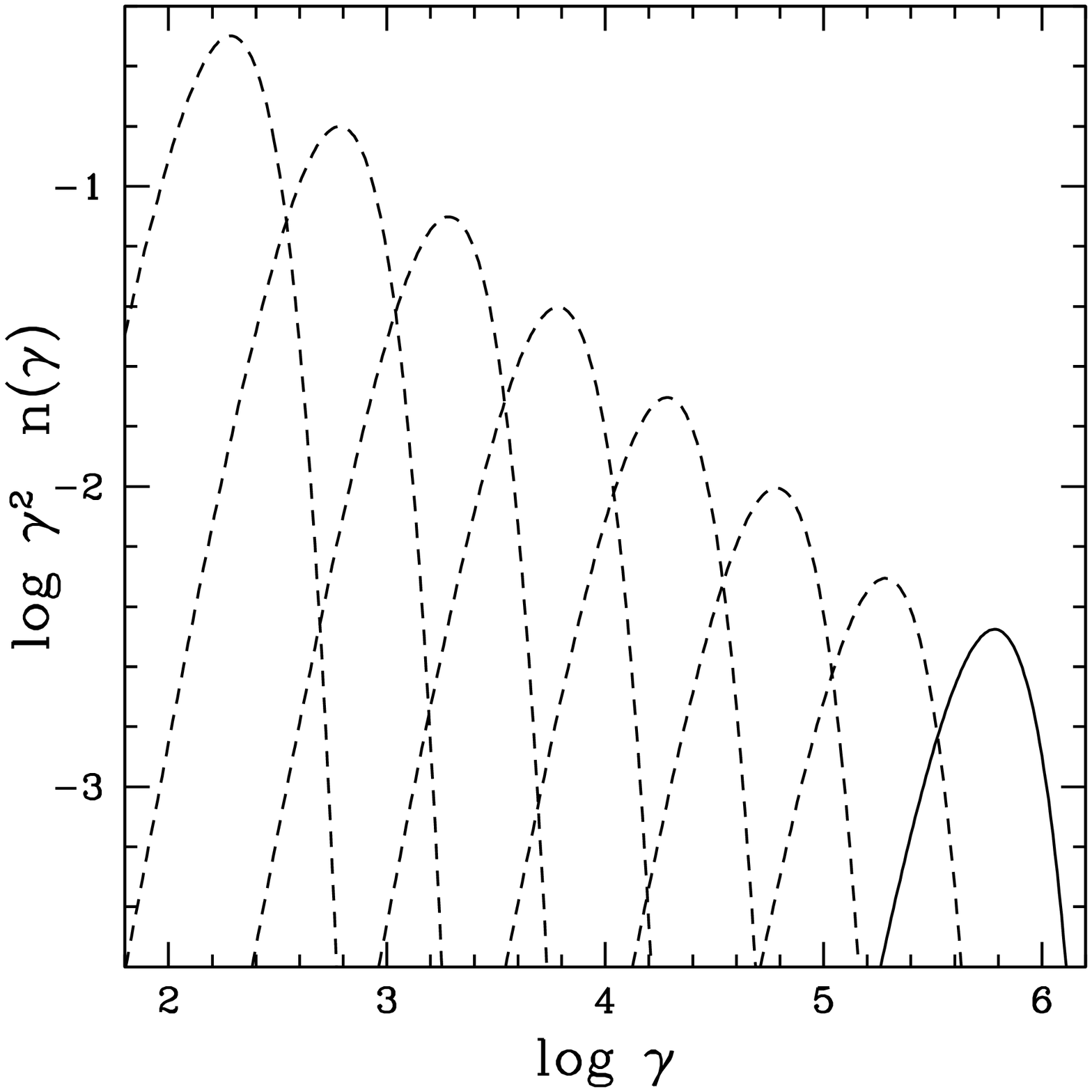}
\caption{Example electron distributions to illustrate discussion
in the text. Left:  uniform and two-part spectra, as discussed in 
Sections \ref{CombineSyICS} and
\ref{AlternativeModels}. The colored sections of the curves 
show the regions actually probed by
observations; the dotted and dashed lines are not constrained by any
 current data.   The continuous curve (a) describes a homogeneous
source with $B = 1 \mu$G throughout; it has
 $n(\g) \propto \g^{-2.4}$, $\g_1 \ltw \g \ltw \g_{br}$;  
$n(\g) \propto \g^{-3.4}$, $\g_{br} \ltw \g \ltw \g_2$;  
and $n(\g) \propto \g^{-4.2}$,  $\g \gtw \g_2$.
The two disjoint curves (b) and (c) describe a two-part model. 
Curve (b), for the high-field region, 
is identical to the lower-energy part of curve (a), 
but shifted in amplitude and energy, to correspond to a $10 \mu$G field;
 it will give exactly the same radio spectrum as curve (a).  
Curve (c), for the low-field region, is identical
to the high-$\g$ part of curve (a), with a small offset for clarity, and
produces the same $\g$-ray emission.  If this
part of the source has a weak magnetic field, $B \ll B_{CMB}$, there will be
no detectable radio emission from electrons in curve (c). Right:  
Example of a sequence of peaked electron spectra, created by 
Alfvenic ISR, in a range of B fields, as discussed in Section
\ref{AlfvenISR}.
 Each individual spectrum here is
$n(\g) \propto \g^2 e^{-(\g / \g_c)^{3/2}}$, an example of
the steady state kernel created
when Alfvenic acceleration  balances radiative losses. The peak
energy $\g_c$, can be found from \eref{AlfvenPeakEnergy} if $W(k) \propto
k^{-3/2}$. The solid line shows one spectrum  with
$\g_c = 10^{5.5}$.  The dotted lines show the same function, repeated for
a range of lower $\g_c$ values.  The spread in $\g_c$ illustrated here
 can be caused by a spread $\sim 125$ in B field.
 The amplitude of each kernel, chosen arbitrarily
here to minic an integrated power law, 
is determined physically by the fractional volume of the source 
at the relevant B field.}
\label{ToyModel}
}
\end{figure}

A simple example reveals the essence of this calculation (also carried out
by A10, Y12).  In curve (a) of the left panel of 
Figure  \ref{ToyModel} I show a broken power-law $n(\g)$,
representative of the
specific $n(\g)$'s chosen by A10, Y12, but easier analytically.  
  Referring back to \eref{ICS_electrons} 
and \eref{synch_electrons}, and using the notation of Figure \ref{ToyModel}, 
this example gives
\be
{ U_{e,sy} \over U_{e,ICS}} = { t-2 \over s-2} 
{ \g_2 \over \g_{br}} \left(
{ \g_2 \over \g_1} \right)^{s-2}
\label{HomogModel}
\ee Once $\g_1, \g_{br}$ and $\g_2$ are chosen, equating the ratio
$U_{e,sy} / U_{e,ICS} $ inferred from the data (as in Table
\ref{OL_energies}) to the ratio calculated from 
\eref{HomogModel} determines the B field required throughout the
source.  The break energies are constrained by observations. The
{\em Fermi}-LAT data require $\g_2 \sim 3.4 \times 10^5$, and the radio
break at 5 GHz requires $\g_{br} \sim 9 \times 10^4 B_{\mu G}^{-1/2}$.
However, $\g_1$ is uncertain, as discussed in
 Section \ref{TotalElectronEnergies}.   It is therefore useful to
carry out this calculation for both high and low $\g_1$ values.

If $\nu_1 = 10$ MHz, Table \ref{OL_energies} gives
$U_{e,sy} / U_{e,ICS} \sim 240 B_{\mu G}^{-3/2}$.  Equating this to
the expression in \eref{HomogModel} gives $B \sim 1.4 \mu$G in the outer
lobes, an electron energy $U_{e,sy} \sim 1
\times 10^{57}$ erg and a magnetic energy $ \sim 7 \times
10^{57}$ erg. The electron spectrum is truncated at $\g_1 \sim
3500$, and the system is field-dominated:  $U_B \sim 7 U_{e,sy}$.  
Alternatively, if  $\g_1 = 10$, a similar calculation gives about the
same field, $B \sim 1.5 \mu$G, but a larger electron energy, 
 $U_{e,sy} \sim 1.5 \times 10^{58}$ erg.  Thus, with this low $\g_1$,
the electron/field balance has reversed.  Because of the abundant
low-energy electrons, the system is now
marginally electron-dominated, with $U_{e,sy} \sim 2 U_B$. This recovers
the  ``quasi-equipartition'' particle-field balance,
similar to results in A10 and Y12.

\subsection{From theory:  inhomogeneous models}
\label{AlternativeModels}

The previous models  treat the radio
lobes as homogeneous systems, with one magnetic field throughout.
Turning to theory, however, we expect the B field to 
have some spatial structure.  Basic physics tells us it is hard to
create a uniform, single-valued B field throughout a complex plasma
structure such as the Cen A lobes.

If the source is inhomogeneous, 
radio emission will come dominantly from plasma in high-B regions,
while the $\g$-rays come from any part of the source which contains
relativistic electrons at the right energies.   This can be illustrated
with a 
two-part spectral model, in which localized, high-B regions (radio-loud)
coexist with an extended, low-B ($\g$-ray-loud) plasma.  The 
electron distributions in the two regions may be totally decoupled;  curves
(b) and (c) of the left panel of
 Figure \ref{ToyModel} show one example.  In this
situation,  the relation 
between $U_{e,sy}$ and $U_{e,ICS}$ (equation \ref{HomogModel}) no
longer applies, so we must abandon the simple result of a quasi-equipartition
field throughout the lobes.  

We do not know what supports the magnetic field in the radio lobes of
Cen A, but MHD theory motivates several possible models.

{\em Turbulent dynamo.}  The outer lobes probably contain 
turbulence, possibly driven by internal velocity shear.  The
turbulence will drive a fluctuation dynamo, supporting a mean 
field, $B_{turb}^2 \ltw 4 \pi \rho v_t^2$.  (Note the field is
dynamically weak if the turbulence is subsonic: $B_{turb}^2 \propto
{\mach}_t^2 p$ where $v_t$ and ${\mach}_t$ are the velocity and
 the Mach number of the
turbulence, and $p$ is the total pressure).
  While the field is spatially intermittent (obeying a
log-normal distribution), simulations suggest that most of the volume
is filled with $B \sim B_{turb}$, and only a small fraction of volume
contains significantly larger fields, especially in the case of
subsonic turbulence (\eg, Schekochihin \etal 2004, Haugen \etal 2004).



\begin{table}[htb]
\caption{\label{OL_Bfields}
The range of  B field  models suggested for  the outer lobes of Cen A.} 
\begin{indented}
\item[]
\begin{tabular}{l c   l  }
\br 
\rule{0pt}{16pt} \hspace{-8pt} 
Method  & B ($\mu$G)  
 & Comments, where discussed
\\[4pt]
\mr
\rule{0pt}{16pt} \hspace{-8pt} 
Minimum radio-loud field$^{\rm a}$ & $ \gtw 0.06 \xi^{2/3}$ 
& Plasma pressure balancing IGM pressure (\S \ref{UsefulLimits})  
\\[4pt]
Field at minimum pressure$^{\rm a}$  & $\sim 0.6 \xi^{2/7}$ 
& Minimum pressure in radio-loud plasma (\S \ref{UsefulLimits})  
\\[4pt]
Homogeneous models$^{\rm a}$  & $\sim 1$
&  Tuned to fit radio and $\g$-ray emission;
\\
& & assumed uniform throughout lobe (\S \ref{CombineSyICS})  
\\[4pt]
Turbulent dynamo$^{\rm b}$  & $\sim (2\!-\!3){\mach}_t$ 
& Magnetic field supported by turbulence;
\\
& & spatial fluctations small if ${\mach}_t \ltw 1$
 (\S \ref{AlternativeModels}) 
\\[4pt]
Dynamic balance field & $\sim 3$ 
& Magnetic field balancing IGM pressure;
\\
& & highest field that can be confined by IGM (\S \ref{AlternativeModels}) 
\\[4pt]
Shock-enhanced fields$^{\rm c}$ & $\sim 10$ 
& Weak shocks throughout source;
\\
& & High B localized to shock regions  (\S \ref{AlternativeModels}) 
\\[4pt]
Self-organized fields$^{\rm c}$ & $\sim1\!-\! 10$ 
& Force-free fields throughout source;
\\
& & Large-scale gradients in B (\S \ref{AlternativeModels}) 
\\[4pt]
\br
\end{tabular}
\item[] $^{\rm a}$ Assumes $\nu_1 = 10$ MHz ($\g_1 \simeq 4000 B_{\mu G}^{-1/2}$)
for low-energy cutoff to electron spectrum. The scaling factor 
$\xi = (1+k)/\phi$ (see \S \ref{UsefulLimits}).
\item[] $^{\rm b}$ ${\mach}_t$ is Mach number  of turbulence, defined in terms
of the adiabatic sound speed;  numerical range for $\Gamma \in (4/3, 5/3)$
and $p = 3.2 \times 10^{-13}$ dyn cm$^{-2}$.
\item[] $^{\rm c}$ Magnetic field value is an estimate only, to
  illustrate an overpressured B field.
\end{indented}
\end{table}

{\em Dynamic balance field.} The largest field that can
be in pressure balance with ambient IGM is   $ B_{dyn}
 \simeq ( 8 \pi p_{IGM})^{1/2} \simeq 2.8
 (n_{-4} T_{keV})^{1/2} \mu$G (using $p_{IGM}$ from Section \ref{IGM_conditions}).
While the lobes are unlikely to contain a uniform field at this level, $B_{dyn}$
is a useful fiducial value.  Stronger
 fields must be either transient or self-confined.

{\em Shocks in transonic turbulence.} Transonic turbulence 
generates a complex network of shocks, in which 
high-field fluctuations can become significant (\eg, Molina \etal 2012).
It is tempting to specuate that the radio-bright filaments (apparent in
Figure \ref{Fig:OuterLobes}) are shocks in transonic
turbulence (\eg, Tregillis \etal 2001). If these are weak shocks
(Mach number of a few), simple compression will boost the
pre-shock field by a factor $\gtw 2$. In addition, there is 
substantial evidence that instabilities set up by  streaming cosmic rays  
enhance the pre-shock magnetic field more
strongly than just the modest compression boost (\eg,
Riquelme and Spitkovsky 2010; Schure \etal 2012). 
 For numerical examples I will
use $\sim 10 \mu$G (roughly ten times higher pressure than lobe plasma
not close to any shock).  

{\em Self-organized magnetic structures.}
Several authors have suggested that self-organized, nearly
force-free fields exist within radio lobes and cavities (\eg,
Lynden-Bell 1996;  Gourgoulias
\etal 2012).  These structures are the end point of
magnetic (or Taylor) relaxation, which develops spontaneously if
magnetic energy and helicity are injected into the system (\eg, Tang 2008,
Benford and Protheroe 2008).  In their simplest versions (\eg, Lynden-Bell
1996, Li \etal 2001), these structures are
characterized by magnetically dominated cores within denser, lower-B 
outer regions.  This model connects well to the frequency-dependent
width of the north lobe (Section \ref{LargeScales}), which could be
caused by a  weaker B field towards the outer edges of the source.
    For numerical examples I will also
use $\sim 10 \mu$G for the high-field regions in this model.

\section{Young models do not work} 
\label{ModelsThatDontWork}

Some authors have suggested the radio lobes are  no older
than the radiative lifetimes of their relativistic electrons.  
In this section I show this cannot be the case. 

\subsection{Radiative lifetimes of relativistic electrons}
\label{RadiativeLifetimes}

Determining radiative lifetimes should be a
straightforward application of
basic physics.  However,  we do not  know the
magnetic field in which a relativistic electron currently sits, let
alone that which has felt over its radiative life.  I therefore
evaluate the maximum possible radiative life, which applies if the particle
has spent most of its life in a weak magnetic field, $B \ll B_{CMB}$.
Details are given in \ref{Appendix:Radiative_life}, where the key
results are \eref{ICS_max} and \eref{sy_max}.

The short lifetime of the
$\g$-ray loud electrons is the most stringent condition on the physics
of the outer lobes. Equation (\ref{ICS_max}) shows
that these electrons (with $\g \sim 3.4 \times 10^5
\!-\! 2.8 \times 10^6$) live {\em no longer than}
 $6.8$ Myr for the lower energies, or
$0.80$ Myr for the higher energies.  These upper limits on the lifetime
hold if  $B \ll 3.2 \mu$G (the CMB equivalent field);  higher
 fields will reduce the radiative lifetimes significantly.  For example, if
$B \sim B_{dyn} \sim 2.8 \mu$G, the lifetime range reduces to $0.47 \!-\! 3.8$
Myr;  overpressure fields (\eg, Table \ref{OL_Bfields}) result in sub-Myr
lifetimes for all $\g$-ray loud energies. 
Thus, if Cen A is older than $\sim 1$ Myr, 
the $\g$-ray-loud  electrons must be undergoing ISR.

The lifetimes of the radio-loud electrons are  also relatively
short.  Because high frequency
observations (H09) suggest a spectral turnover around $\sim 5$ 
GHz, the maximum lifetime of electrons radiating 
 at that frequency is
of interest.  From \eref{sy_max}, the longest possible synchrotron
lifetime is $t_{sy}^{max} (5 \mbox{GHz}) \simeq 25
B_{now, \mu G}^{1/2}$ Myr, where $B_{now}$ is the field in which the
electrons currently sit. The largest plausible value of 
$B_{now}$ might be  $\sim 10 \mu$G in the radio-loud regions
(as in Section \ref{AlternativeModels}  and Table \ref{OL_Bfields}).
Even this optimistic estimate  would push $t_{sy}^{max}(5 \mbox{GHz})$ 
only to  $\sim 80$ Myr.   If Cen A is older than this, 
the radio-loud electrons in the outer lobes must also be undergoing ISR.

\subsection{The outer lobes can't be as young as they look}
\label{OlderThanTheyLook}

It is easy to show that Cen A must be
older than the radiative lifetime of its $\g$-ray-loud
electrons.  For instance, if  Cen A is  1 Myr old,
the ends of the outer lobes must have propagated from the
AGN at $\simeq 0.99 \mu^{-1} c$.  Even if the lobes lie close to 
the sky plane, this violates observations: 
there is no sign of relativistic beaming or
sidedness on these large scales. In addition (referring to models
I will present in Section \ref{OL_drivenflow}),  near-lightspeed propagation
of the outer ends would require a relativistic internal flow speed (which
also  violates observations), and  a very low ambient density,
(which seems unlikely given Cen A's central position within
its group of galaxies (as in Section \ref{IGM_conditions}). 

Alternatively, various authors have argued that Cen A is no older than the
synchrotron age for radio-loud electrons in the outer lobes.  H09
estimated $t_{sy} \sim 30$ Myr for electrons radiating at
5 GHz; they and several authors since have
taken this as the true age of the source. 
 Y12 made a similar argument, but chose
 $t_{sy} \sim 80$ Myr for electrons radiating at 1 GHz.
These ages still violate observations.  
I will show (in Section \ref{OL_drivenflow}) that  growing the radio source so
quickly requires  highly supersonic flows within the outer
lobes.  Such flows would be easy to detect, especially their
terminal shocks which would be seen as hot spots.

\subsection{Ages based on spectral steepening}
\label{Spectral_ages}

 If the dynamic age of a radio galaxy is not known, the
 radio and/or $\g$-ray spectrum of the source is often  used to derive 
a ``spectral age'' (\eg, H09 or Y12 for Cen A).  These
calculations make the critical assumption that the relativistic
  electrons seen in radio or $\g$-rays are energized only at their
  initial insertion into the system, after which they undergo only
  radiative losses.  In that situation, the energy  at which the electron
spectrum steepens is directly related to the elapsed time since the
process started, through  \eref{ICS_full} and \eref{sy_full}.
However, this calculation is only a good estimate of the age of the source
if we know that no other physics has affected the electrons.
  Because several possible ISR mechanisms can alter
the electron distribution (referring to models I will discuss in
Section  \ref{InSituReacceleration}), I argue that spectral ages are
not robust unless backed up by independent evidence, for instance from
dynamic models of the source's evolution.

It is worth noting that
some  spectral-aging models include a variant on the required ISR. 
These models (\eg, H09, Y12) 
assume that new electrons are continually ``injected'' into the radio 
lobes as the pre-existing electrons cool.  The injection
mechanism is not specified, but is typically thought of as new particles
supplied from the AGN.  In  Cen A, those electrons
would have to travel at nearly lightspeed in order 
to offset the rapid  losses of the existing $\g$-ray-loud
population in the outer lobes.
   Such fast streaming is difficult given the finite 
plasma density and disordered magnetic field likely to exist in the lobes.  
One might argue instead that the new electrons are injected
locally, pulled from a thermal population by shock or turbulent
acceleration (as in 
Section  \ref{InSituReacceleration}).  This is possible,
but if such processes exist, they would also provide ISR for 
the existing relativistic electrons, thus modifying the energy spectrum
of those electrons and undermining any simple relation between spectral break 
energy and source age.

\section{Older models which may work}
\label{ModelsThatWork}

In the previous section I argued that the radiative lifetimes of the
relativistic electrons in Cen A are  shorter than any plausible
dynamical age of the radio lobes.  In this section I justify this
assertion with  three simple models for growth
of the outer lobes:  a  buoyant plume, a magnetic tower, 
and a flow-driven model. 
 Two timescales  set the stage for what follows.  

{\em The sound-crossing time in the IGM}. Using the adiabatic sound speed, 
($c_s^2 = \Gamma k T /  {\bar m}$ with adiabatic index $\Gamma = 5/3$ and 
mean particle mass ${\bar m}
= 0.6 m_p$) and recalling the length of the source is 
 $D \sim 300/\mu$ kpc, the sound crossing time over this distance
in the IGM is 
\be
\tau_{sound} = {D \over c_{s,IGM}} \sim {570 \over  \mu 
 T_{keV}^{1/2} } ~~ {\rm  Myr}
\label{soundcrossing_age}
\ee
I will show in this section that $v_{end} \ltw
c_{s,IGM}$ for the two slower models (buoyant plumes and magnetic
towers), 
so than Cen A must be at least as old as $\tau_{sound}$.  In 
a momentum-driven flow, I will show that the age of the source can
be pushed down to perhaps half of $\tau_{sound}$, but no younger.

{\em The time needed to excavate the outer lobes.} 
 The enthalpy content of an outer lobe is
$ \Gamma p V / (\Gamma-1) \sim 1.8 \times 10^{59}/\mu$ erg
if the lobe plasma is internally subrelativistic ($\Gamma = 5/3$), 
or $\sim 2.9 \times 10^{59}/\mu$ erg if the plasma is
internally relativistic ($\Gamma = 4/3$). 
Scaling the time-averaged power  from the AGN 
to its current strength ($P_{AGN} =  10^{43}P_{43}$ erg/s 
to each side of radio source; Section \ref{InnerDrivers}), we can estimate
the time needed to excavate the outer lobes:
\be
\tau_{exc} = { \Gamma \over \Gamma -1} { p V \over P_{AGN}}
\sim  {570 \!-\! 910 \over \mu P_{43}}  ~~ {\rm  Myr}
\ee
This expression assumes the plasma within the lobe is dominated by thermal
energy;  the numerical range is for $\Gamma \in ( 4/3 , 5/3)$.  If the lobe contains significant transonic flows or dynamically strong
magnetic fields the excavation time will be longer than these values.

\begin{table}[htb]
\caption{\label{Model_summary}
Summary of  dynamic models for outer lobes;  see text for details.
$\mu$ is the
cosine of the angle between the radio lobes and the sky plane. }
\begin{indented}
\item[]
\begin{tabular}{l c  c  c}
\br 
\rule{0pt}{16pt} \hspace{-8pt} 
 & Buoyant plume & Magnetic tower & Flow-driven lobe
\\[4pt]
\mr 
\rule{0pt}{16pt} \hspace{-8pt} 
Advance speed & $v_{end} < c_{s,IGM}$& $v_{end} \sim c_{s,IGM}$ 
 & $v_{end} > c_{s,IGM}$
\\[4pt]
Age  & $ \gtw  570 / \mu $ Myr  & $ \sim 570 / \mu$ Myr 
& $\sim (300\!-\!600)/ \mu$ Myr 
\\[4pt]
Necessary conditions
 &  internally subsonic  & high-B plasma core 
     & internally transonic
\\[4pt]
Magnetic structure & turbulent dynamo, & large-scale order, 
& high-B (shocks), 
\\
& small-scale fluctuations & large-scale gradients & low-B (intershock)
\\[4pt]
ISR mechanism & Alfven turbulence & reconnection & shock acceleration
\\
&  & $+$ Alfven turbulence & $+$ Alfven turbulence
\\[4pt]
\br 
\end{tabular}
\end{indented}
\end{table}

The three models to be described
in this section are closely connected to the different
magnetic field models (described in Section \ref{InternalConditions}),
and also to the  competing ISR mechanisms (to be described in \S
\ref{InSituReacceleration}). To help the reader keep it all straight, in
Table \ref{Model_summary} I summarize the three models including their
probable connections to B field structure and ISR mechanisms.

\subsection{Slower growth:  buoyant plumes}
\label{SlowGrowth}

The plasma at the outer end of a mostly-hydrodynamic radio lobe
  feels two forces,  buoyancy (asuming the lobe is less dense than the
ambient IGM), and ram pressure (from
fluid coming in behind it).  The slowest growth happens when 
buoyancy drives the lobe. We know, of course,
that the outer lobes of  Cen A do not resemble the turbulent, entraining plumes
which some authors (\eg, Bicknell 1984, DeYoung 1986)  have
suggested as models for tailed radio galaxies. The 
outer lobes do not show the 
 gradual broadening, going away from the source, which is 
characteristic of such plumes.  Nor is there any sign 
of the lateral spreading expected if the end of a plume has reached the
neutral buoyancy level in the surrounding IGM. 
However, the growth of the  outer lobes may still be dominated by
bouyancy if the buoyant force exceeds ram pressure force  at the
end of the lobe.

There is not much work in the fluid literature describing the advance
speed of the end of a driven buoyant plume, but we can estimate an
upper limit.  We know that the speed at which a small buoyant bubble
rises through an atmosphere is determined by the balance of buoyant
and drag forces.  If the ambient medium is hydrostatic, in the gravity
of the galaxy group, it is easy to show that this buoyant rise speed
is no larger than the sound speed in the external medium, and very
likely much less.  We can speculate that a similar limit applies to
the buoyancy-dominated end of a continuous plume, so that the end
advances into the IGM at $v_{end} \ltw c_{s,IGM}$.  Thus, the age of
a bouyancy-driven radio lobe will be at least as long as $\tau_{sound}$, and 
probably longer.

The internal state of a buoyancy-driven radio lobe is probably close to the
homogeneous models of Section  \ref{CombineSyICS}.  Nothing in the model
calls for large-scale spatial gradients.  Some level of MHD turbulence
is likely, and indeed is necessary for ISR (as will be shown
in Section \ref{InSituReacceleration});  but if the turbulence is
subsonic, it does not induce significant inhomogeneities (Section 
\ref {AlternativeModels})

\subsection{Intermediate growth: magnetic towers}
\label{MagTowers}

An alternative  model obtains when 
the AGN power creating the radio lobes is dominated by the magnetic
field.  Although some plasma flow can be included in 
a classic magnetic tower, the energy is mostly carried by Poynting flux
(Lynden-Bell 1996, Li \etal 2001).  Numerical models
of magnetic towers propagating into an ambient IGM  (\eg,   Nakamura \etal 2006)
show a magnetically dominated interior, with
the field close to a force-free configuration. 
As the tower grows, it comes into lateral pressure
balance with the IGM;  hoop stresses concentrate the field toward the axis
and maintain lateral collimation, while allowing vertical growth
as the front end advances into the IGM (\eg, Uzdensky and MacFadyen 2006).
Both analytic models and simulations show the end advances at
 $v_{end} \sim c_{s,IGM}$ into the IGM (Lynden-Bell 1996, Nakamura \etal 2006). 
The growth rate of other self-organized models, such as those suggested
by Tang (2008) or Benford and Protheroe (2008) has not been worked out, but
we might guess that such structures develop similarly to magnetic towers.

Because the internal state of a magnetically self-organized radio lobe 
includes large-scale gradients in both the B field strength and the plasma
density, these models are 
among the inhomogeneous-field models discussed in Section  
\ref{AlternativeModels}.

\subsection{Faster growth: flow-driven lobes}
\label{OL_drivenflow}

If organized plasma flow from the AGN continues to the end of the
outer lobes, ram pressure can dominate the growth of the lobe and grow
the source faster than bouyancy or magnetic pressure. 
A simple model can illustrate this idea, which I summarize
here;  details are given in \ref{Appendix:Toy_Model}. 

Assume a channel flowing within each radio lobe carries 
all of the energy and momentum to drive the lobe. 
At the outer end of the  lobe -- which advances less rapidly than
the speed of the plasma within the channel -- that plasma
must slow down and move aside.  
The channel will therefore
 be surrounded with a ``cocoon''
which acts as a reservoir for mass and energy
from the channel.  This cocoon is the observed radio lobe.  This model
derives, of course, from standard models of FRII radio sources, but 
must be modified to agree with the lack of strong outer hot spots in FRI
sources. The model has two important parts, as follows. 

(1) The
advance speed of the end, $v_{end}$, is governed by momentum conservation,
where the outer end of the channel impacts the IGM.  
 The momentum flux carried by the channel will be spread over an area ($A_{end}$)
at least as large as the area of the channel ($A_c$),
 and possibly larger.  If the channel impacts
 the end of the lobe  with density $\rho_c$ and speed $v_c$, $v_{end}$ is
governed by $ \rho_{IGM} v_{end}^2 \ltw \rho_c v_c^2$
(\ref{WS_advance_speed_short}, \ref{Limiting_forms}).
If the lobe and IGM are also in pressure balance, 
and the end propagates supersonically into the IGM, 
the channel flow must be {\it internally} transonic:
$\Gamma_c \mach_c^2 \gtw \Gamma_{IGM} \mach_{end}^2$
(\ref{Mach_number_relation}, \ref{Limiting_forms}).
The lower limits on $v_c$ and ${\mach}_c$  
 obtain when $A_{end} \to A_c$;  if the momentum flux is spread
over a larger impact area, the advance speed $v_{end}$ is correspondingly
lower.    

(2) The flow power in the channel, $P_c$, can be found directly from
energy conservation, if we account for growth of the lobe volume and
work done on the IGM.  This calculation depends on the aspect ratio $a/D$ of
the lobe:  if the lobe grows self-similarly, with $a/D$ remaining constant,
more power is required to grow the lobe at a given $v_{end}$  than if
the lobe radius $a$ remains constant.  I parameterize this uncertainty with
a factor $1 \ltw f_V \ltw 3$, describing the 
range from constant-radius growth to self-similar growth. The general
result for $P_c$, \eref{General_channel_power},
contains two terms:  one describing  
 the  growth of the full lobe, and an additional factor
describing the extra power needed to maintain the
channel flow.  As with the advance speed, the result $P_c$
depends on the working surface area,
$A_{end}$.  The minimum $P_c$ needed to grow the source, \eref{Power_result},
 occurs when $A_{end} \to A_c$;  for larger $A_{end}$ values, higher
power is needed.

\begin{table}[htb]
\caption{\label{Flow_driven_models}
Flow-driven models for outer lobes.
Models assume lobe length $300/ \mu$ kpc,
lobe radius $a = 90$kpc, and pressure $p =  3.2 \times 10^{-13}$\dyncm2.
$f_V \in (1,3)$ describes the geometry of the expanding lobe.
 $\mu$ is the cosine of the angle
between the radio lobe and the sky plane.    $r_c/a$ is
the ratio of channel radius to lobe  radius.  The minimum value of $P_c$,
from \eref{Power_result},  is
reached when momentum flux is exerted only over the area of 
the channel; a larger working area requires a larger $P_c$. 
}
\begin{indented}
\item[]
\begin{tabular}{c c c c c c  l }
\br 
\rule{0pt}{16pt} \hspace{-8pt} 
Age & $ v_{end}$ &  ${\mach}_{end}$$^{\rm a}$  &  ${\mach}_c$$^{\rm a}$ 
    & Power carried in channel$^{\rm b}$  & comments
\\
(Myr) & (km s$^{-1}$) &  &   & (erg s$^{-1}$)
\\[4pt]
\mr 
\rule{0pt}{16pt} \hspace{-8pt} 
$30/\mu$  & 9800  & 25. & $\gtw 25.$
& $P_c \gtw 1.9 f_V \times 10^{44} \left[ 1 + 600 r_c^2/a^2 \right]$
&  unlikely
\\[4pt]
$100/\mu$  & 2900   & 7.5 & $\gtw 7.5$
& $P_c \gtw 5.6  f_V \times 10^{43} \left[ 1 + 56 r_c^2/a^2 \right]$
&   unlikely 
\\[4pt]
$300/\mu$  & 980   & 2.5 & $\gtw 2.5$ 
& $P_c \gtw  1.9 f_V  \times 10^{43} \left[ 1 + 6.3 r_c^2/a^2 \right]$
&  challenging 
\\[4pt]
$600 / \mu$ & 480 &  1.2  & $\gtw 1.2$
& $P_c \gtw 9.3 f_V \times 10^{42} \left[ 1 + 1.4 r_c^2/a^2 \right]$ 
& possible 
\\[4pt]
\br 
\end{tabular}
\item[] $^{\rm a}$ 
The internal Mach number  of the channel
is $\mach_c =  v_c / c_{s,c}$;  the Mach number of
the end relative to the IGM is  $\mach_{end} = v_{end} / c_{s,IGM}$.
\item[] $^{\rm b}$
Values for $P_c$ assume $\Gamma_c = \Gamma_{IGM} = 5/3$;  if $\Gamma_c =4/3$, 
the leading term in $P_c$ would increase by 1.6;  the second
term  inside brackets would decrease by 0.6.
\end{indented}
\end{table}

To apply this analysis, we can choose a source age, $\tau$, and find
the necessary end speed ($v_{end} = D / \tau$). Because we are 
searching for models in which the source is
 younger than $\tau_{sound}$, we 
require supersonic propagation of the end into the IGM:
$v_{end} \gtw  c_{s,IGM}$.
I keep $r_c/a$ as a free parameter, but note that spectral tomography
of some FRI sources (Katz-Stone and Rudnick 1996, Katz-Stone \etal 1999)
suggests that $r_c$ is not too small a fraction of the source radius
(perhaps $r_c/a \sim 1/3 - 1/2$).  Putting these choices into
\eref{Power_result}, along with values of the lobe radius $a$ and
ambient pressure $p_{IGM}$, gives the power needed to grow the source
at rate $v_{end}$. 
Table \ref{Flow_driven_models} presents a range of choices for
$v_{end}$, with the associated source ages, Mach numbers
 and required core powers.  By keeping two criteria in mind, we
immediately see that the younger  models cannot work for  Cen A.  (1) Cen A 
does not have bright outer hot spots.  This alone rules
out the younger models, because Mach numbers ${\mach}_c \gtw 2 \!-\! 3$ will 
create a strong shock at the outer end;  such a shock
 would be highly visible as a radio-loud hot spot.  
(2) Younger models require a much higher core power in the past, 
averaged over the age of the source, than the power now being put out
by the AGN ($\sim 10^{43}$erg/s, Section \ref{InnerDrivers}).
 This is especially true if $r_c \ltw a$,
but holds even if $r_c \ll a$ (which would suggest a
very small, very bright outer hot spot).  Of course it may be that the 
AGN in Cen A is currently running at much lower power than it has
over most of its active life -- but such models are unattractive. 

The internal state of a flow-driven lobe will very likely be turbulent,
driven for instance by internal shear flows (also see Tregillis \etal 2001).
Because the internal flows in the most plausible models are 
transonic with  ${\mach}_c \sim 1-2$, the turbulence they drive will
include weak shocks throughout the radio lobes.  Thus, these models
are included in the inhomogeneous models discussed in Section  
\ref{AlternativeModels}.

\subsection{Compare dynamic models to the data}
\label{Compare_dynamic_models}

The simple models presented so far have been general, and might 
apply to any FRI-type of radio galaxy.  How well do they do when
confronted specifically with the data?  
Key observational results on Cen A are as follows. 

\begin{itemize}

\item{\em Morphology.}
Both lobes have a generally uniform
brightness distribution.  There is no sign of hot spots or any surface
brightness enhancement towards the outer ends. 
The outer lobes are not limb brightened, 
but rather center filled (as shown by slices through the image  
in Figure \ref{Fig:OuterLobes}).  Their brightness distribution is 
not totally smooth, however. Some large-scale features ($\sim 30-60$ kpc)
exist in each lobe,  including the 
Northern Loop (Junkes \etal 1993) and a low surface  brightness 
``hole'' at the outer end of the north lobe;  a similar region on the
southeast edge of the south lobe; and the vertex/vortex structure
in the  southern lobe  (Feain \etal 2011).

\item
{\em Limb darkening.} As mentioned in Section \ref{LargeScales}, 
the size of each lobe depends on the observing frequency. The north 
lobe is significantly broader in $\g$-rays than in radio (Y12). 
In radio, the full length of both lobes is detected only below 
$\sim 10$ GHz (H09); the 
full radio width of the north lobe is seen only
below $\sim 1$ GHz (apparent by comparing low-frequency radio
images, from McKinley \etal 2013
and Stefan \etal 2013, to the 5 GHz image from Junkes \etal 1993). 
Without the $\g$-ray data, this would normally be interpreted as 
spectral aging (plasma in the extremities of the lobes is older and the
electron spectrum has steepened there;  \cf\,
 Section \ref{Spectral_ages}).  However, because we know that
ISR is needed for the both the $\g$-ray-loud and radio-loud electrons,
passive spectral aging is not the answer.   Large-scale gradients in
the B field and/or the ISR mechanism must exist in the radio lobes.

\item
{\em Polarization.} Both lobes show strong, ordered, large-scale
 polarization. Parts of both lobes are polarized up to 30-50\% on
 scales $\sim 60-70$ kpc
(Junkes \etal 1993).  As with other radio galaxies, the
polarization vectors in Cen A tend to correlate with local geometry.
The component of the B field in the sky plane tends to lie
 along bright ridges or filaments or along the outer edges of the
radio lobes. 
These data strongly constrain any model of small-scale, 
isotropic  magnetic turbulence within lobes, 
because  any more than a few random magnetic ``cells''
 along the line of sight would 
significantly depolarize the signal (\eg, Burn 1966).

\item
{\em Filaments.}  The outer lobes are filled with
 striking filamentary structures apparent in 
Figure \ref{Fig:OuterLobes}.  The filaments  are not sparse;
   they have a covering factor close to unity. 
The lack of limb 
brightening rules out edge effects (\eg, surface instabilities) as the
cause of the filaments;  they must be internal to the lobes.
 Measurement of a few bright filaments 
reveals a typical  width of a few kpc and a
typical  emissivity enhancement of a factor of $\sim 10$
(relative to the mean lobe emissivity, assuming cylindrical
filaments). 

\end{itemize}

\noindent
How do the
three dynamic models fare against these data?

\begin{itemize}

\item
{\em Buoyant models}.  Because this  model is to first-order
internally homogeneous, it is consistent with uniform, center-filled
lobes.   However, just because of that homogeneity, the model
is  severely challenged by the 
large-scale polarized structures, frequency-dependent limb darkening and
high-contrast radio filaments.  Additional physics will be needed
to explain these phenomena -- for instance the effects of a subsonic flow
entering the lobe from the AGN.

\item
{\em Magnetic towers.}  Because this model invokes system-scale
gradients in the B field,  it is easily compatible with large-scale
polarized structures as well as the observed limb darkening.
Simple axisymmetric versions of the model 
(\eg, Nakamura \etal 2006) fail because they predict a radio-faint central
cavity that is not seen.  Spatially more 
complex models may avoid this problem, and may also 
provide a possible explanation
of the filaments (\eg, Tang 2008).  However, 
 these models have not yet been developed to where they can
be robustly tested against the data.

\item {\em Flow-driven models.}  Because this model is intrinsically
  two-part (a channel flow within a surrounding cocoon), it agrees
  well with center-filled emission and has the potential to explain
  {\it lateral} limb darkening as well (if physical conditions in the
  cocoon are sufficiently different from those in the channel).
  Internal shocks driven by transonic flow provide an attractive
  model of the filaments, and may exist on large-enough scales
to account for the large-scale polarized features.
 The model is seriously challenged,
  however, by the lack of any sign of channel deceleration or
  spreading towards the outer ends of the lobes.

\end{itemize}

This comparison shows that the internally homogeneous, buoyant model
does the poorest job of matching the data.  The magnetic tower and
flow-driven models do better, but objections can be made to 
simple versions of each one.  Furthermore, any successful dynamic
model of Cen A must incorporate one or more ISR mechanisms;  that is
the topic of the next section.
In Table \ref{Model_critique} I summarize
how well, or badly, each of the dynamic models compare to the data.

\begin{table}[htb]
\caption{\label{Model_critique}
Critique of models suggested for Cen A.  Buoyant model assumes Alfvenic
ISR; magnetic tower assumes reconnection and Alfvenic ISR; flow-driven
model assumes shock and Alfvenic ISR. }
\begin{indented}
\item[]
\begin{tabular}{l c  c  c}
\br 
\rule{0pt}{16pt} \hspace{-8pt} 
 & Buoyant plume & Magnetic tower & Flow-driven lobe
\\[4pt]
\mr 
\rule{0pt}{16pt} \hspace{-8pt} 
Center-filled lobes & easy & possible$^{\rm a}$
 & easy
\\[4pt]
Large-scale polarization & hard & easy & possible$^{\rm a}$
\\[4pt]
Lateral limb darkening &  hard & easy & needs work$^{\rm b}$
\\[4pt]
Outer end limb darkening & hard & needs work$^{\rm b}$ &  hard
\\[4pt]
Radio-bright filaments & challenging & needs work$^{\rm b}$ & easy
\\[4pt]
Radio-$\g$-ray power ratio & challenging & easy & easy 
\\[4pt]
Steep $\g$-ray  spectrum & needs work$^{\rm b}$ & needs work$^{\rm b}$  
& needs work$^{\rm b}$  
\\[4pt]
Power-law radio spectrum & challenging &  probable$^{\rm c}$  & easy
\\[4pt]
\br 
\end{tabular}
\item[] $^{\rm a}$ ``Possible'' means the model satisfies the criterion if
conditions are right within the source (large-scale shocks in flow-driven 
model;  high-order modes excited in magnetic tower models.)
\item[] $^{\rm b}$ ``Needs work'' means that relevant models not yet 
sufficiently developed to meet the data (nature of the cocoon for
flow-driven models; distribution of radio emissivity 
in magnetic tower models;  high-energy end of particle spectrum in  ISR models).
\item[] $^{\rm c}$ ``Probable'' means model works if current theories of
reconnection ISR are valid.
\end{indented}
\end{table}

\section{In situ reacceleration in the outer lobes}
\label{InSituReacceleration}

In the previous section I showed that the outer lobes of Cen A
are  much older than radiative lifetime of the relativistic electrons which
make the lobes shine in radio and $\g$-rays. This proves that ISR is
required throughout the outer lobes.  
In this section I discuss how that might happen. I begin by demonstrating
that diffusion is likely to be slow within the radio lobes, so that 
local (in situ) reacceleration is needed.  I then discuss three
competing ISR models -- Alfven wave turbulence, shocks and reconnection --
in the specific context of dyanmic models I have introduced for Cen A.
  I argue that Alfvenic ISR is probably the most likely mechanism in
all three models, but that it is not yet well enough understood
to confront the data fully.  I also argue that shock and reconnection ISR
are very attractive alternatives {\em if} the necessary conditions exist
within the outer lobes.

As a prelude to this discussion, note that all three ISR models are
sensitive to particle scattering on microphysical scales (possibly as
small as the gyroradius, $ r_L = \g m c^2 / e B$).  ISR models
generally assume this scattering exists, with some mean free path
$\lambda_{mfp}$.  Unfortunately, $\lambda_{mfp}$ is not well
understood. Some authors assume the optimistic Bohm limit, in which
$\lambda_{mfp} \sim r_L$, but detailed modelling suggests the
Bohm limit is not often reached, and scattering on a spectrum of MHD
waves may be better described by   quasi-linear (QL)
theory (\eg, Casse \etal 2003; Dosch \etal 2011).  

In the usual QL 
approach, Alfven wave turbulence is assumed to exist throughout the
source.  Let it have a spectrum $W(k)$, 
normalized so that the turbulent energy density is $u_{turb} = \int W(k) dk$.  
Relativistic particles  at energy $\g$ interact most strongly
with waves at their resonant wavenumber, $k_{res}(\g) \simeq e B / \g m c^2$
(equivalently, with resonant wavelength
$\lambda_{res} \simeq r_L$). If $W(k)$ decreases with $k$ above
some driving scale $k_{min} = 2 \pi / \lambda_{max}$, we can write
$u_{turb} \sim k_{min} W(k_{min})$ for the total turbulent energy, and
$u_{res}(\g) \sim k_{res}(\g) W[k_{res}(\g)]$ for the energy in waves at or above
the resonant wavenumber. Because the ratio $u_{res}(\g)/u_{turb}$ is
critical in what follows, we must develop numerical estimates, as follows.

To estimate $\lambda_{max}$ in Cen A, we recall 
fluid flows, where $\lambda_{max}$ is commonly taken as some large fraction of
the channel width.  In Cen A, we might also use the loops and filaments
in the radio image to estimate $\lambda_{max}$;  inspection of the image
in Figure \ref{Fig:OuterLobes} shows their scale is also some large
fraction of the lobe width.  Based on these arguments, I will scale
$\lambda_{max}$ to 30 kpc.  For comparison, the resonant wavelength of
$\g \sim (0.34-2.8) \times 10^6$ electrons seen in $\g$-rays is
$r_L(\g) \simeq (0.7\!-\!5.6) \times 10^{15} / B_{\mu G}$
cm, so that $k_{res}(\g) \gg k_{max}$.

We also need the ratio  $u_{res}(\g)/u_{turb}$, which measures
the fraction of turbulent energy in resonant waves.  
It is traditional in the literature to represent  the wave spectrum 
as a power law, $W(k) \propto k^{-r}$, with $r \in (1,2) $ typically. If
this assumption is correct, 
 $u_{res}(\g)/u_{turb} = \left(k_{res}(\g)/k_{min} \right)^{-(r-1)} \ll 1$;
only  a very small fraction of the total turbulent energy is available
to interact resonantly with the high-$\g$ particles we can observe. 
This $k^{-r}$ spectrum is commonly invoked in turbulence studies
(\eg, Zhou \etal 2004 and references therein)
and found in numerical simulations  which assume a low-$k$ driver  
(\eg, Schekochihin \etal 2004). 
However, these results may not be the last word.  
Bicknell \& Melrose (1982) speculated that strong coupling between
Alfven and magnetosonic waves will subject Alfven waves to the strong damping
that magnetosonic waves suffer and terminate the cascade at wavenumbers
well below $k_{res}(\g)$.  In addition, modern work such as Brunetti \etal
(2004) considers more general possibilities for wave generation, and thus
more general wave spectra than the simple $k^{-r}$ power law.  In this
section I will use this simple power law for numerical examples, but the
reader should be aware that the real situation may be more complicated.

\subsection{Need for distributed reacceleration}
\label{DiffusionStuff}

It is easy to show that ISR sites must be distributed throughout the outer
lobes.  
If the B field within the lobes is turbulent,
propagation away from ISR sites will be by diffusion, as the particles
scatter on MHD turbulence.\footnote{If the B
field is ordered, as in magnetic tower models, the particles can stream 
along field lines
at the Alfven speed, but much slower cross-field propagation is diffusive.} 
If tangled field lines extend 
 throughout the lobes, particles will diffuse along the field lines at a rate
 described by a coefficient $\kappa_{\parallel} \simeq c \lambda_{mfp} / 3$. 
As discussed above, some authors assume Bohm
diffusion applies, with $\kappa_{\parallel} = \kappa_B \simeq c
r_L / 3$. This is the slowest diffusion likely for this geometry.
Alternatively, other authors (\eg, Casse \etal 2003) find the true
behavior is closer to the coefficient predicted by QL
theory, 
\be
\kappa_{\parallel} \simeq { 24 \over \pi \eta } \kappa_B
 { u_{turb} \over u_{res}(\g) }
 \label{QL_diffusion}
\ee
where $\eta = u_{turb} / u_B$ is the ratio of
turbulent to total magnetic energy. 

For $k_{res} \gg k_{min}$, and  a wave spectrum $W(k)$ which decreases with
$k$,  clearly $\kappa_{\parallel} \gg \kappa_B$.
However, we are lucky here:  despite the large uncertainty in the correct
value of $\kappa_{\parallel}$, we can show that 
the distance a relativistic electron travels from its origin is small.
Recall that any ISR mechanism must be able to offset the 
most rapid radiative losses, with lifetimes
only a few Myr for the $\g \gtw 
10^6$ electrons which radiate at the high end of the $\g$-ray band.  
For example,  picking $B = 1 \mu$G and $\g = 10^6$ and assuming
QL scattering, the diffusion distance in 1 Myr is only  
$\sim \left( 4 \kappa_{\parallel} t_{rad} \right)^{1/2} \sim 16$ kpc 
for $\g$-ray loud electrons.\footnote{ This example
used $B = 1 \mu$G,$B_{rad} = 3 \mu$G, $\eta = 1$,
$r = 5/3$ for a Kolmogorov spectrum
and $\lambda_{max} \sim 30$ kpc. The diffusion coefficients are then
  $\kappa_B \simeq 2 \times 10^{25}$cm$^2$ s$^{-1}$ and $\kappa_{\parallel}
\simeq 2.0 \times 10^{31}$cm$^2$ s$^{-1}$.}
If Bohm diffusion rules, the diffusion distance
is much smaller, less than a kpc.  (Different
choices for $B, \lambda_{max}$ and the turbulent exponent $r$ do not affect
this qualitative conclusion.)
Thus -- because the diffusion distance is so small --
 we see that ISR sites must 
exist throughout the outer lobes.

\subsection{Alfven turbulent acceleration}
\label{AlfvenISR}

This ISR mechanism is commonly invoked for diffuse nonthermal
plasmas, including radio galaxies in general (\eg, Lacombe 1977, Eilek
1979), and Cen A in particular (H09, O'Sullivan \etal
2009, Wykes \etal 2013).  
Relativistic particles scatter on resonant waves
 and slowly diffuse in momentum space, in a
second-order ISR process.

\paragraph{Acceleration rates.}
Two conditions must be met in order for Alfvenic acceleration to be a
useful model for  Cen A.   The first condition is speed:
the acceleration time, $t_{acc}$, must be no longer than the radiative
loss time, $t_{rad}$ (from \ref{GeneralLifeTime}).  
To determine $t_{acc}$, we can follow standard QL theory 
(\eg, Lacombe 1977) to determine the momentum diffusion 
coefficient,\footnote{This expression approximates
 equation 9 of Lacombe (1977), with $p = \g m c$;  note the exact numerical 
coefficient of $D_p$ depends on details of the wave field directionality,
which vary author to author.} 
\be
D_p \simeq { \pi^2 e^2 v_A^2 \over c^3} W\left[k_{res}(p) \right]
\label{QL_accn}
\ee
This expression is, of course, closely related to the QL diffusion coefficient
\eref{QL_diffusion}, because both processes depend on resonant particle
scattering by Alfven waves.
 From \eref{QL_accn} we can  estimate the acceleration time,
\be
t_{acc}(\g) \simeq { p^2 \over D_p} \simeq { 8 \over \pi} { \g m c \over e  B}
{ 1 \over \eta \beta_A^2} \left[ { u_{turb} \over u_{res}(\g)} \right]
\label{AccTimeAlfven}
\ee
This form shows that acceleration is faster for higher $\eta = u_{turb}/u_B$, 
higher $\beta_A = v_A/c$,
and higher fractional resonant wave energy, $u_{res}/u_{turb}$. 
To illustrate,
we can  assume $\eta = 1$, and again use simple power law wave spectrum,
$W(k) \propto k^{-r}$, with $r = 5/3$, $\lambda_{max} = 30$ kpc.
Using these in \eref{AccTimeAlfven}, with  $\g = 10^6$,
we find $t_{acc}(\g) \ltw 1$ Myr if $v_A \gtw 4100$ km s$^{-1}$. 
This is a key result:  because Alfvenic acceleration is slow, $v_A$ must
be high if Alfvenic turbulence is to keep the radio lobes of Cen A shining.

\paragraph{Problems with the simple model.}
Speed isn't everything, however.  A model of ISR must also produce an
 electron energy distribution, $n(\g)$, which corresponds to observed 
synchrotron
and ICS spectra. {\it This is a major problem for this simple model.} 
 In the context of the  homogeneous-field model  
(Section  \ref{CombineSyICS}, also curve (a) of the left panel of 
Figure \ref{ToyModel}),  $n(\g)$ must be a power law at lower
energies ($\g \ltw 10^5$ in a $\mu$G field), and steepen strongly
 into and through the ICS range.  However,
in a steady state, particle-conserving situation, turbulent Alfvenic
 acceleration
in the presence of radiative losses creates an electron
energy distribution which is {\it peaked} at the energy where $t_{rad}(\g)
\simeq t_{acc}(\g)$:
\be
\g_{c,A}^2 \simeq { 3 \pi^2 \over 4} { e \over \sigma_T} { B \over B_{net}^2}
\eta \beta_A^2 \left[ { u_{res}(\g) \over u_{turb}} \right]
\label{AlfvenPeakEnergy}
\ee
(Borovsky and Eilek 1986, Stawarz and Petrosian 2008). 
Electrons on the high-energy side of such a
 peaked distribution may explain the steep $\g$-ray spectrum,
but cannot explain the broadband, power-law radio spectrum 
(as in the left  panel of Figure \ref{ToyModel}). Thus the simple 
model must be extended if Alfvenic ISR is to explain the full
spectrum from Cen A. 

One  extension is to a spatially variable
magnetic field, say from large-scale gradients in the radio lobes.
A range of B fields will broaden the effective electron energy
distribution (as in the right panel of Figure \ref{ToyModel}).  This
broad $n(\g)$ will in turn  distribute the synchrotron
kernel function (which peaks at $\nu_{sy} = a B \gamma_c^2(B)$) in 
frequency, creating a broader, possibly power-law
radio spectrum (Eilek and Arendt 1986). Another extension of the model lets
the wave spectrum vary in time and space, deviating from the usual power law.
Brunetti \etal (2004) present a specific example which results in
 a broad, slowly varying  $n(\g)$ distribution.
Finally, a third extension invokes localized injection of low-energy
electrons throughout 
the radio lobe, enabling Alfvenic ISR to create a broader 
electron $n (\g)$ (\eg, Borovsky and Eilek 1986).
This injection could be provided by nearby 
shocks or reconnection sites, both described in the next subsection.

\subsection{Additional ISR mechanisms}

While Alfvenic ISR is the  mechanism most commonly invoked for the 
diffuse radio lobes of Cen A, shock acceleration and reconnection 
acceleration are two attractive alternatives if conditions within
the source enable these mechanisms to work.

\paragraph{Shock acceleration.}

Although not often considered for ISR in diffuse radio galaxy lobes,
shock acceleration is  commonly invoked to explain galactic
cosmic ray acceleration in supernova remnants, and ISR in radio galaxy
jets and hot spots.  
It is well known that particles scattering on
self-generated turbulence on either side of a shock are accelerated
into a power law energy distribution which agrees well with observed
synchrotron spectra (\eg, Drury 1983).  Because this
first-order process is fast, we can envision a shock as a localized
``injection event'', very quickly boosting pre-shock particles to high
energies.

We must, however, be aware of a major uncertainty in shock
theory, namely the scattering mean free path, $\lambda_{mfp}$ (usually
expressed in terms of the shock diffusion coefficient, $\kappa_s \sim
c \lambda_{mfp}/3$). As mentioned in Section  \ref{DiffusionStuff}, 
some authors assume Bohm diffusion (the best case 
for rapid shock acceleration, \eg, Kang 2011),
 but others assume QL scattering (\eg, Giacalone 2005). 
We should also note that $\kappa_s$, close to the shock, is not necessarily
the same as $\kappa_{\parallel}$ elsewhere in the diffuse radio lobe; 
the turbulence is likely to be quite different in the two regions.

Keeping this uncertainty in mind, we can nonetheless
check the energy range attainable,
and compare it to the apparent steepening around $\sim 5$ GHz in the radio
spectrum of Cen A.  The timescale for diffusive shock acceleration
is $t_{acc}(\g) \simeq F(s) \kappa_s / v_s^2$, where $v_s$ is shock speed
and $F(s) = 3 s (s+1)/(s-1)$ depends on the shock compression ratio $s$
(\eg, Drury 1983).  
Shock acceleration in the presence of radiative losses will create a
power law electron $n(\g)$, up to the energy at which
 $t_{rad}(\g_c)\simeq t_{acc}(\g_c)$:
\be 
\g_{c,s}^2
\simeq { 18 \pi \over F(s)} { \kappa_B \over \kappa_s} { e \over \sigma_T} 
{   v_s^2 \over c^2} { B \over B_{net}^2} 
\ee 
To illustrate, let the physical field be $10 \mu$G, and 
choose $s = 2$ for a weak shock. 
Scaling $v_s$ to $10^3$ km s$^{-1}$, and $\kappa_s$ to $\kappa_B$, we get 
$\g_{c,s}\simeq 1.5 \times 10^7  v_{s,3} 
\left( { \kappa_B /  \kappa_s} \right)^{1/2} $.
Comparing to Figure \ref{ToyModel}, we see that -- if Bohm diffusion
holds -- $\g_{c,s}$ is high, well above $\g_{br} \ltw 10^5$ needed for
the 5-GHz spectral break in the high-B region.  We can speculate on
two possible ways to reconcile this estimate for $\g_{c,s}$ with the
data.  Because radiative losses are rapid for these high $\g$'s, 
electrons at $\g \sim \g_{c,s}$  will lose most of their energy very
quickly.  Noting that radio observations cannot resolve the small
postshock regions, the observed spectral break may mean we are seeing most
of the electrons a few Myr after they've been shocked.  Alternatively,
Bohm diffusion may be too optimistic.  If  $\kappa_s \gg
\kappa_B$ we could be seeing the intrinsic high-$\g$ cutoff of the
power law created by shock acceleration.

\paragraph{Reconnection acceleration}

The theory of  reconnection ISR is still a work in progress;
even the basic physical picture is not yet agreed on.  We can explore
this mechanism with a simple, two-dimensional reconnection model, as
follows.  Let two regions of oppositely directed B field be separated
by a thin dissipation layer. The plasma on each side of the layer
flows in, at some $v_{in}$, inducing an electric field ${\cal E}
\simeq v_{in} B / c$.  Magnetic energy dissipated in this reconnection
layer goes to plasma heating and/or acceleration of suprathermal
particles. Particles within this region are subject to two different
types of energization -- parallel and perpendicular.

One ISR possibility  is parallel energization within the reconnection
layer.  This occurs if  the particle moves 
unimpeded through some (large) distance, 
$\lambda_{\parallel}$, along the inductive potential drop created by
the inflow (\eg, Litvinenko 1999,
Romanova and Lovelace 1992, Benford and Protheroe 2008).  
The key requirement here is that $\lambda_{\parallel}$  be large (compared
to microscales such as the electron Larmor radius);  this
is the opposite limit to strong, Bohm-like scattering. In this situation,
the particle gains energy $\Delta E \simeq e {\cal E} \lambda_{\parallel}$,
or 
\be
\g_{\parallel} \simeq { e B \over m c^2} \beta_{in} \lambda_{\parallel}
\label{Parallel_recn}
\ee
where $\beta_{in} = v_{in}/c$.  Detailed modelling shows that reconnection
events form substantial nonthermal tails, 
for $\g \ltw \g_{\parallel}$.  These tails
are often interpreted as power laws, but this is still an active 
research area (\eg, Kagan \etal 2013 and references therein).
As a simple example, we can
guess $B \sim 1 \mu$G, and $\lambda_{\parallel} \sim 1$ kpc. 
If we also take $v_{in} \sim 300$ km s$^{-1}$ (on the order of 10\% of $v_A$
suggested in Section \ref{AlfvenISR}), we get $\gamma_{\parallel} 
\sim 2 \times 10^9$, easily high enough to keep Cen A shining.
However, both $\lambda_{\parallel}$ and 
$v_{in}$ are very sensitive to local conditions;  I return to this issue
later in this section.

The other ISR possibility is perpendicular energization across the reconnection
layer.  If the plasma in the region is sufficiently turbulent, particles will
scatter back and forth in the converging inflow, gaining energy in a
first-order Fermi process (\eg, 
de Giouveia dal Pino and Lazarian 2005, also Drury 2012).  
Just as with shock acceleration, the key step here is the strength of
the turbulent scattering, measured by $\lambda_{mfp}$. 
The acceleration time given by Drury (2012) can be approximated, for our
simple reconnection-layer picture, as  
$t_{acc} \sim 3 \lambda_{mfp}/v_{in}$.\footnote{Bosch-Ramon (2012) gives an even
faster estimate for $t_{acc}$, pushing the perpendicular scattering rate close 
to the Bohm limit. If this is the case, $\g_{c,rec}$ will be even higher
than in my illustration here.} 
The power-law $n(\g)$ created in this process will extend
up to an energy at which $t_{acc}(\g_c) \simeq t_{rad}(\g_c)$, or
\be
\g_{c,rec}^2 \simeq  2 \pi {r_L \over \lambda_{mfp}} 
{e \over \sigma_T} \beta_{in}  { B \over B_{net}^2} 
\label{Perp_recn}
\ee
Illustrating again with a $1 \mu$G field,  $B_{net}^2 = 8 \pi ( u_B +
u_{CMB})$, $\beta_{in} \sim 10^{-3}$ and guessing
 $\lambda_{mfp} \sim 100 r_L$ we find
$\g_{c,rec} \sim 6 \times 10^{7}$, again high enough to keep 
Cen A shining.

We must remember that three important parameters in 
\eref{Parallel_recn} and \eref{Perp_recn} are  sensitive to local
conditions.   One is $\lambda_{\parallel}$, the length of the 
reconnection layer. 
If reconnection is driven by MHD turbulence, $\lambda_{\parallel}$ is 
limited above by the  coherence length of
the turbulence (\eg, Servido \etal 2011, Zhdankin \etal 2013).  
Alternatively, if the radio lobes
contain self-organized magnetic structures, $\lambda_{\parallel}$ will
be larger, up to 
a significant fraction of the domain size (\eg, Benford and Protheroe
2008).  A second parameter is the
  inflow speed, $v_{in}$, which is a matter of active current debate.
Early work suggested $v_{in} \ll v_A$ in quiescent systems (\eg, 
Zweibel and Yamada 2009 and references therein). 
 However, a large body of modern
work points to much faster reconnection in collisionless
and/or turbulent systems, typically reaching $v_{in} \gtw 0.1 v_A$
 (\eg, Shay \etal 2004, Markidis \etal 2012, Kowal
\etal 2012).   Furthermore, large-scale motions in
the system can drive reconnection at inflow speeds comparable to the
bulk motion speed (\eg, Yamada \etal 1997;  Gekelman \etal 2012).
A third parameter is $\lambda_{mfp}$, the scattering mean free path. 
As with spatial diffusion and shock acceleration, $\lambda_{mfp}$
must be assumed {\em ad hoc}, and
can vary from Bohm to QL scattering, with
 orders of magnitude inbetween.  

To summarize, the examples here show that reconnection  ISR is an attractive
alternative to shock acceleration, and has the potential
to keep Cen A shining {\em if local conditions in the reconnection regions
are favorable.}  However, many details of the physics remain to be worked
out before either version of reconnection ISR can be tested robustly 
against the data. 

\subsection{Compare the models to the data}
\label{Compare_ISR_models}

The summary of ISR models in this section has mostly been  general.  How
do they compare to the data on Cen A?  Key results germane to
ISR are as follows.

\begin{itemize}

\item {\em Short lifetimes.}  A successful ISR model must be able
to energize relativistic electrons rapidly enough to offset radiative
losses.  For Cen A, the most stringent constraint is the $\sim 1$ Myr
lifetime for $\g \gtw 10^6$ electrons observed in $\g$-rays.

\item {\em Radio spectrum.}
The integrated radio spectrum of the outer lobes of Cen A is a power law,
with spectral index $\alpha_{rad} \sim 0.7$, from $\sim 10$ MHz to 
$\gtw 10$ GHz (Alvarez \etal 2000;  H09). 
 The narrowband radio spectrum is relatively constant within 
the lobes, showing some fluctuations but no large-scale
trends (Combi and Romero 1997, McKinley \etal 2013).\footnote{Problems 
with incomplete flux recovery make the
apparent gradients  in the spectral index map based on PAPER data
between 123 and 173 MHz (Stefan 
\etal 2013) questionable.}  

\item{\em Gamma-ray spectrum.}
The integrated $\g$-ray spectrum is steeper than the radio spectrum.
It has been fit by a power law between $\sim 200$ MeV to $\sim 7$ GeV
(A10, Y12), but a slowly curving spectrum also fits the data.
 The photon statistics are not yet good
enough to detect spectral gradients within the lobe.

\item{\em Ratio of $\g$-ray to radio power.}
As discussed in Section \ref{TotalElectronEnergies}, the observed $\g$-ray
and radio powers require a specific distribution function for 
relativistic electrons {\em if} the source is homogeneous.  This constraint is
relaxed if the radio and $\g$-ray emission come from different regions
(for instance, the radio from high-B regions and the $\g$-rays from 
throughout the lobes).

\item{\em Connection to dynamic models.}  
An ISR model must make sense in the context of a successful dynamic
model.  For instance, shock acceleration can only apply if transonic
flows exist within the source; reconnection acceleration can apply
only if  magnetic field reversals exist within the source.

\end{itemize}

\noindent
How do the three ISR models fare against these data and constraints?

\begin{itemize}

\item
 {\em  Alfven wave  acceleration} is very likely happening in the 
radio lobes of Cen A. Because Alfven waves are easy to generate and
hard to damp, a field of Alfven waves is likely to
exist throughout the outer lobes, whether they obey the flow-driven 
or magnetically-driven models.  If the plasma is tenuous enough,
Alfvenic ISR can  offset radiative losses.
 However, current models of Alfven
wave acceleration are challenged by the broadband, power-law
  radio spectrum; 
if Alfvenic ISR is the sole ISR mechanism, more physics is
needed.  Possible extensions of the models include spatial gradients in the B
field and/or the turbulence structure, and a dynamic model 
of the Alfven wave spectrum.

\item
 {\em Shock acceleration} will be a key part of the flow-driven  models 
of Section  \ref{OL_drivenflow}.  If these models apply to Cen A, weak shocks 
will  exist here and there throughout the outer lobes.
  Standard shock acceleration
theory -- even allowing for uncertainties in the detailed physics --
suggest that even weak shocks can easily maintain a power-law
energy spectrum for radio-loud electrons.  Thus, if transonic flows 
exist within the lobes of Cen A, shock acceleration is a very attractive
ISR mechanism.  Because the shocks will be localized, high-B regions,
they probably coexist with Alfvenic ISR operating elsewhere in the radio
lobes.

\item
{\em Reconnection acceleration} almost certainly exists at some level
within the outer lobes of Cen A. 
On small scales, reconnection is an intrinsic part of MHD
turbulence (\eg, Retin\`o \etal 2007, Servido \etal 2011).
If Cen A obeys the flow-driven model,
small-scale reconnection ISR should be operating throughout the outer
lobes, probably augmenting Alfvenic ISR from the same turbulent regions.  
On large scales, reconnection is essential for the magnetic relaxation 
involved in magnetic tower models (\eg, Tang and Boozer 2004,
 Benford and Protheroe 2008).  If these models describe Cen A, 
 reconnection ISR may play the major role in maintaining
radio emission from the high-B regions, coexisting
with Alfvenic ISR operating elsewhere in the source.

\end{itemize}

\subsection{Constraints on plasma density} 

As a final note to this ISR discussion, we must return to
Alfvenic ISR.  I have argued that this mechanism is 
operating throughout the lobes of Cen A, no matter which
dynamic model holds.  However,
the condition required for for rapid Alfvenic ISR
is critical, and possibly contentious.  In order for this mechanism to
be effective, the Alfven speed must be high,  
$v_A = B / (4 \pi n_p m_p )^{1/2}  \gtw 4000$ km s$^{-1}$ (derived
in Section \ref{AlfvenISR} under the assumption of a 
broad-band $k^{-5/3}$ wave spectrum). 
If a proton-electron plasma carries the Alfven waves, this high speed
requires a low density, $n_p \ltw 3 \times 10^{-7} B_{\mu G}^2$ \cm3,
which appears inconsistent with two recent papers (Stawarz \etal
2013, O'Sullivan \etal 2013).  Both papers suggested much higher densities 
in the lobes, $n_p \sim 10^{-4}$\cm3.  Such   high density, 
if it exists, would be comparable to that in the surrounding IGM, and
would reduce $v_A$ to $\sim 200 B_{\mu G}$ km s$^{-1}$,
slow enough that Alfvenic acceleration would fail as an ISR mechanism in
Cen A.  However, the 
 data in both papers are subject to more than one interpretation.

Stawarz \etal (2013) reported a {\it tentative} (their italics) detection
of  diffuse, thermal X-ray emission at $\sim 0.5$ keV
from small regions coincident with 
 the southern radio lobe of Cen A.  Assuming this emission comes from plasma
within the lobe, they derive a density $\sim  10^{-4}$\cm3;  but
it seems equally likely that this emission is related to the Galactic 
foreground (for instance part of the large-scale structure detected by ROSAT,
shown in Figure 7 of Stawarz \etal), and/or to the surrounding
IGM of the galaxy group in which NGC 5128 resides.

O'Sullivan \etal (2013) detected a Faraday rotation signal, $\sim 10$
\radm2, apparently associated with inner parts of north and south
outer lobes in Cen A.  They suggest the Faraday rotation does not
come from the IGM of the galaxy group or from a skin associated with
the lobe/IGM interface, but rather comes from
plasma within the outer lobes.   Their particular model 
(turbulent depolarization on 20-kpc scales and a 
sub-$\mu$G B field taken from A10), also  requires $n \sim 10^{-4}$\cm3 
  within the radio lobes.

However, the situation regarding Faraday rotation associated
with radio galaxies is complex, and the final story may not yet be
known.  A large body of previous work has used
the lack of detected internal Faraday rotation to place upper limits 
on thermal gas density with RG lobes (\eg,
 Spangler and Sakurai 1985, Laing and Bridle 1987, Garrington
and Conway 1991;  also Feain \etal 2009 for Cen A).   These limits are
consistent with the plasma in the lobes
 being much less dense than  surroundings.  
In addition, as Wykes \etal (2013) note, the
 existence of X-ray cavities in galaxy
clusters coincident with radio lobes and tails (\eg,  Wilson \etal 2006 for
Cyg A, and Wise \etal 2007 for Hydra A)  also points to the plasma density 
within the lobes being significantly lower than that in
 the surrounding IGM.   

It is possible, of
course, that Cen A is the exception to the general rule, and is as dense
as its surroundings.   But it has long been known
that Faraday rotation in many radio galaxies seems to be associated
with the immediate surroundings of the radio lobes, either at the lobe/IGM
interface or in disturbed IGM close to the radio source  (\eg,
Dreher \etal 1987;  Rudnick and
Blundell 2003;  Guidetti \etal 2011). While this connection is 
not yet understood -- we have no good model of how a Faraday-strong shell
forms around the radio lobes --  Cen A may be just another example of
this effect.

\section{Discussion and conclusions}
\label{Conclusions}

In this section I summarize my previous arguments in this paper, discuss
which of the models I have proposed might best describe Cen A, and speculate
on how these models connect to other evidence we have on the recent history
of the galaxy.

\subsection{Summary I:  tentative  models for Cen A}

In this paper I have shown that the radio galaxy Cen A is on the order of
1 Gyr old;  substantially younger models violate observations.  Because this
is much older than the radiative lifetimes of the relativistic electrons
seen in radio and $\g$-rays, in situ reacceration (ISR) must be ongoing
throughout the 300-kpc-scale radio lobes. 
 I developed the argument in three parts, as follows.

$\bullet$ I showed (in Section \ref{InternalConditions}, also 
Table \ref{OL_Bfields})
that homogeneous, quasi-equipartion ($u_B \sim u_e$) models
are {\it not} required  by the data. Inhomogeneous models, in which high-B
and low-B regions coexist, are also consistent with the data and probably
closer to the real physics.  The high-field regions could be shocks 
within the outer lobes or force-free regions within a 
large-scale magnetic structure.

$\bullet$ I introduced three toy models to describe the dynamical
growth of Cen A (in Section \ref{ModelsThatWork}, also Tables \ref{Model_summary}
and \ref{Model_critique}).  
(1) The slowest-growing model assumes the outer ends
of the radio lobes are driven  by {\it buoyancy}.  (2) A second
model assumes the radio lobes are magnetically dominated, with
large-scale organized structures akin to {\it magnetic towers}.  (3) A
third, faster-growing model assumes radio lobe growth is driven by
{\it internal plasma flow}.  These models require Cen A to be 
at least several hundred Myr old, and possibly older than a Gyr.  By
comparison, the radiative lifetimes range from a few Myr for
$\g$-ray-loud electrons, to at most $\sim 80$ Myr for radio-loud
electrons; thus ISR is needed.

$\bullet$ I explored (in Section \ref{InSituReacceleration}, also 
Table \ref{Model_critique})
three competing ISR models relevant to Cen A.
  (1) {\it Alfven wave turbulence} can explain the data
if the radio lobe plasma  is tenuous and 
 highly variable in time and/or space.  
(2)  {\it Weak shocks} 
are an  attractive ISR mechanism in the transonic, flow-driven model;
they may account for most or all of the radio emission in that model. 
(3)  {\it Reconnection} is another attractive 
ISR mechanism, especially in the magnetically driven model,
  {\it if} reconnection site physics is consistent with  current  thinking.

\subsection{Summary II: which models might work for Cen A?}

Having introduced three dynamic models, and explored what ISR mechanism(s)
can work within each one, I then compared the models to the
data (in Section \ref{Compare_dynamic_models}
and Section \ref{Compare_ISR_models};  also 
Table \ref{Model_critique}).  While the simplest form of
the buoyant model had severe challenges, two models remain viable:
flow driven and magnetically driven.

\begin{itemize}

\item  {\em Flow-driven models} are the closest to traditional models
of FRII sources, and may provide a good description of many tailed
FRI sources. 
 Their two-part structure (channel flow plus surrounding cocoon) gives
a natural explanation for the transverse spectral structure seen in Cen A.
If the channel flow is transonic, the internal shocks it drives
are an attractive  way to explain the radio filaments and 
radio spectrum of Cen A.

\item {\em Magnetically driven models}, while less popular in the 
radio galaxy literature,  are an intriguing alternative to
flow-driven models.  
The large-scale, self-organized  structures inherent
in these models are a tempting explanation for the 
large structures and gradients seen in total and polarized radio emission.
If these models describe Cen A, large-scale reconnection sites probably
supplement Alfvenic ISR throughout the lobes.  
 
\end{itemize}

However, neither of these models has been developed 
to the point where it can confidently be said to describe Cen A. Until my
suggested scenarios in this paper are supported by dedicated modelling and/or
simulations, they remain no more than speculations. While the internal
physics of competing ISR mechanisms clearly needs further work, the most 
immediate -- and addressible -- test of each model  is under what conditions
it can explain the large-scale structure of Cen A.

Flow-driven models are perhaps most 
challenged by the lack of brightening at the outer
ends of the lobes, where the channel flow should be decelerating and/or
developing terminal shocks.  I have speculated that lower Mach number flows
will not suffer such problems, but -- because most work on flow-driven
models has considered high Mach number flows -- it is not yet clear
under what conditions, if at all,  models driven by slower flows 
can avoid disruptive instabilities
and create large-scale radio lobes resembling those in Cen A.

Magnetically-driven models are perhaps most challenged by the lack of
large, synchrotron-faint internal cavities as predicted by existing
models.  I have suggested that the radio lobes may contain more complex,
self-organized magnetic structures, but this remains to be demonstrated.
Work is especially needed to learn
whether AGN-driven plasmas can access the necessary high-order
magnetic modes in a soft-boundary system (such as a radio lobe in the IGM),
and whether the plasma/field mix inside such systems is 
sufficiently radio-loud to create radio lobes resembling those in Cen A.

\subsection{Connect  to the recent history of the galaxy}
\label{RecentHistory}

To close, some possible connections between Cen A and recent events
in the history of its parent galaxy deserve mention.

\paragraph{Very recent history:  disrupted  jets?}

Although active jets from the AGN are feeding the kpc-scale inner
radio lobes, no detectable jets connect the inner lobes to the outer lobes on 
either side of the AGN (Neff \etal 2014). If ongoing energy
transport requires a detectable jet, then both outer lobes are
disconnected from their engine.  Are they dying?  If so,
how quickly will they fade from view?

Because the radio- and $\g$-ray-loud electrons require frequent ISR in
order to keep shining, we must ask, on what timescale the ISR will
cease if energy flow to the radio lobes is disrupted.  Turbulence is
the common denominator in all three ISR mechanisms.  It is clearly
required for Alfven-wave ISR, and it is implicated in our two
alternative ISR models.  Thus, the lobes will fade when the
turbulence decays.

We know that nondriven fluid turbulence will decay in no more than
several turnover times, $\tau_t = \lambda_t/v_t$ (where
$\lambda_t$ and $v_t$ are the characteristic scale and velocity of the
turbulence. We can estimate $\lambda_t \sim \lambda_{max} \sim 30$ kpc from the
images,  but we have no direct probe of
$v_t$.  The flow-driven model allows a possible {\it ansatz}:  if
the field is 
maintained by a turbulent dynamo,  $B^2 / 4 \pi \ltw \rho
v_t^2$, and thus $v_t \gtw v_A$.  Alternatively, with the magnetically 
driven model we can directly take $v_A$ as the turbulent turnover speed.
Thus, our  limit $v_A \gtw 4000$ km/s (from Section \ref{AlfvenISR}) suggests
$\tau_t \ltw 8$ Myr.  We therefore expect turbulence in the
outer lobes to decay in no more than $\sim 30$ Myr. This is longer than
the lifetime of the $\g$-ray-loud electrons, and  comparable to the
lifetime of the radio-loud electrons;  the lobes will fade quickly 
if the turbulence stops being driven.

It follows that, if the outer lobes are disconnected from the AGN
right now, this situation cannot have lasted long. Whatever disrupted
the energy flow from the AGN must have happened no more than a few tens
of Myr ago. Did the AGN go through a
quiescent period, for the past few tens of Myr, and then revive itself only
a few Myr ago?  If so, was this part of a regular duty cycle, or is it 
evidence of an unusual disruptive event in the galaxy's recent history?

\paragraph{Previous history:  did a merger launch the radio galaxy?}

We have seen that Cen A is  on the order of a Gyr old. It is
interesting to compare this age range to the dynamic history of NGC 5128.  
The evidence suggests that at least two significant
events -- encounters or megers -- 
have  happened to this galaxy in the past few Gyr.

One  line of evidence comes from the dramatic gas/dust disk in
the core of NGC5128. Because it is
 warped and twisted  we must be
seeing it fairly soon after its formation, as it settles into a stable
state within the galaxy's gravitational potential. The exact rate
at which the disk settles depends on details of 
the galaxy's structure.  If the galaxy is prolate, models suggest
the disk is only $\sim 200$ Myr old (Quillen \etal 1993);  if 
the galaxy is oblate, the disk is $\sim 700$ Myr old (Sparke 1996).
If, as generally assumed, the disk has been left behind by a recent
merger, that event happened less than a Gyr ago.

Another line of evidence  comes from the galaxy's stars; 
NGC5128 has a complex star formation history.  While the bulk of 
the stars are at least $10$ Gyr old, recent work has found a younger population,
which may be as young as $\sim 2\!-\!3$ Gyr (\eg, integrated 
light spectroscopy of
globular clusters, Woodley \etal 2010; resolved studies of
halo stars, Rejkuba \etal 2011).  If this recent burst of star formation 
was triggered by an encounter or merger, that event happened
$\sim 2\!-\!3$ Gyr ago.

While we do not know what initiates an activity cycle in an AGN,
the ages of these two recent events are intriguing. 
The (relatively young) age of the disk  and the (somewhat older)
age of the young stellar population
in NGC5128 bracket the ages that the dynamic models predict for
the outer lobes. 
Did one of these recent encounters also trigger
AGN activity and create the radio galaxy Cen A?

\ack

It is a pleasure to thank Susan Neff,  Frazer Owen, Sarka Wykes, Hui Li
and Martin Hardcastle for many 
 illuminating conversations about radio galaxy physics in general, and 
Cen A in particular.  Insightful comments from the referees contributed
significantly to this paper.  
 Norbert Junkes kindly provided his Parkes image of Cen A.
This research made use of the NASA/IPAC Extragalactic Database (NED) which is 
operated by the Jet Propulsion Laboratory, California Institute of Technology,
under contract with the National Aeronautics and Space Administration.
The National Radio Astronomy
Observatory is a facility of the National Science Foundation operated
under cooperative agreement by Associated Universities Inc.


\begin{appendix} 

\section{Electron radiative lifetimes}
\label{Appendix:Radiative_life}

If we do not know the historical magnetic field through which an electron has
propagated, we cannot uniquely determine its age. In this
Appendix I set up limiting cases which can be useful in the absence of
such historical knowledge. 

The basic  equations 
\eref{Single_particle_power_ICS} and \eref{Single_particle_power_synch}
(in Section  \ref
{TotalElectronEnergies}) for radiative losses by a single particle, due
to synchrotron radiation and inverse Compton scattering (ICS), allow a useful
definition of the particle's radiative lifetime, $t_{rad}$: 
\be 
A \gamma \int_0^{t_{rad}} \left( u_{rad}(t) + u_B(t) \right) dt
\simeq A \gamma \langle u_{rad} + u_B \rangle t_{rad} = 1 
\label{GeneralLifeTime}
\ee 
Here,  $A = (4/3) c \sigma_T$,  $u_{rad}$ and $u_B$ are the energy
densities in local radiation and magnetic fields, 
and the angle brackets define the
time-averaged radiation and magnetic fields seen by the 
particle historically.

For ICS, 
 we know the particle $\gamma$ directly, and we can write 
\eref{GeneralLifeTime} as
\be
t_{ICS}(\gamma) = { 1 \over A \gamma \langle u_{rad} + u_B \rangle }
\label{ICS_full}
\ee
In many applications $u_{rad}$ is that of the CMB 
and does not vary with
time.  In this case, the longest possible life for an ICS electron happens
when $u_B \ll u_{CMB}$ throughout the particle's life, giving
\be 
 t_{ICS}^{max}(\gamma) = { 1 \over A
  \gamma \langle u_{CMB} \rangle } \simeq { 2.3 \times 10^6 
  \over \gamma } ~ {\rm Myr}
\label{ICS_max}
\ee

However, for synchrotron radiation at a given $\nu_{obs}$,
 we only know the particle $\g$ as a function of the magnetic field
{\it in which the particle now radiates}, $B_{now}$:
 $\nu_{obs} = a \gamma^2 B_{now}$.  The  equivalent to \eref{ICS_full} is 
\be
t_{sy}(\nu_{obs}) = \left( { a \over \nu_{obs}} \right)^{1/2} 
{  B_{now}^{1/2}
\over A \langle u_{CMB} + u_B \rangle }
\label{sy_full}
\ee
In this situation we can again find a limiting case by assuming 
$u_B \ll u_{CMB}$ throughout the particle's life:
\be
t_{sy}^{max} (\nu_{obs}) = \left( { a \over \nu_{obs}} \right)^{1/2} 
{ 8 \pi B_{now}^{1/2} \over A \langle B_{CMB}^2\rangle } 
\simeq 57 { B_{now, \mu G}^{1/2} \over \nu_{GHz}^{1/2}} ~ {\rm Myr}
\label{sy_max}
\ee

\section{Momentum-driven growth of a radio lobe}
\label{Appendix:Toy_Model}

In this appendix I set up a simple model of flow-driven growth for the
outer lobes (or tails)
of an FRI radio source such as Cen A. The model is motivated
by successful models of FRII sources,  and also by observations 
(\eg,  Katz-Stone \& Rudnick 1996, Katz-Stone \etal 1999) 
which reveal complex spectral structure within the tails of some FRI sources.
Consider a two-part radio lobe or tail: 
energy and momentum is carried along a central
channel, with radius $r_c$,  cross-section $A_c = \pi r_c^2$,
density $\rho_c$ and speed $v_c$.  The
channel is surrounded by a larger and nearly static cocoon with  radius
$a$, which we observe as the radio lobe.   Both lobe and channel have
length $D$. The channel, the cocoon and the ambient IGM are all
in pressure balance at some pressure $p$. 
All flows are subrelativistic and have no more than a
dynamically small magnetic field.  

 \paragraph{Growth rate of the lobe.}
The front end of the channel, which is also the front end of the lobe,
 advances into the IGM at speed $v_{end} = dD/dt$.
  The advance rate depends on the
area, $A_{end}$,  over which the momentum flux from the channel 
is spread when it reaches the
contact surface between the lobe and the IGM. Momentum conservation in the
frame of the contact surface gives
\be
 \rho_c \left( v_c - v_{end} \right)^2 + p
= \left( \rho_{IGM} v_{end}^2 + p \right) { A_{end} \over A_c}
\label{WS_advance_speed_full}
\ee
(\eg, Norman \etal 1983, or Mizuta \etal 2004).
It is easy to show from this relation
 that  $v_{end} \ll v_c$ if $\rho_c \ll \rho_{IGM}$;  thus 
\be
 \rho_c  v_c^2 + p \simeq
 \left( \rho_{IGM} v_{end}^2 + p \right) { A_{end} \over A_c}
\label{WS_advance_speed_short}
\ee
This expression can be rewritten in terms of the Mach
numbers\footnote{
 The internal Mach number  of the core
is $\mach_c =  v_c / c_{s,c}$;  the Mach number of
the end relative to the IGM is  $\mach_{end} = v_{end} / c_{s,IGM}$.}, as
\be
\Gamma_c {\mach}_c^2 = \Gamma_{IGM} {\mach}_{end}^2{ A_{end} \over A_c}
+ \left( { A_{end} \over A_c} -1 \right)
\label{Mach_number_relation}
\ee
Unfortunately, the 
area $A_{end}$ is not well specified by current models or data.  It 
 may be as small as the cross section of the channel, $A_c$, 
or it may be larger, up to the full cross section of the lobe, $\pi a^2$. 
We thus have useful inequalities:
\be
\rho_c v_c^2 \gtw \rho_{IGM} v_{end}^2 ~; \qquad 
\Gamma_c {\mach}_c^2 \gtw \Gamma_{IGM} {\mach}_{end}^2
\label{Limiting_forms}
\ee
The lower limits are reached when $A_{end} \to  A_c$.

\paragraph{Power needed to grow the lobe.}

Energy usage in the lobe depends on the evolution of its geometry as it
grows, which is also not well specified by models or data.  In one extreme
case, the lobe radius may stay constant while the outer end advances.
Alternatively, and 
perhaps more likely, the lobe may grow 
self-similarly, keeping the ratios $r_c/a$ and
 $a/D$ constant.  I parameterize this
uncertainty as 
$ dV/dt = 3 \pi a^2 f_V v_{end}$, where $f_V \in (1,3)$ describes the
range from constant-radius growth to self-similar growth.

The power carried in the channel flow is 
\be
P_c = \pi r_c^2  v_c \left[ {\Gamma_c p \over \Gamma_c -1}  + 
{1 \over 2 } \rho_c v_c^2 \right]
\label{jet_pressure_high_beta}
\ee
Our goal is to write $P_c$ in terms of known quantities.
To do this, we can use the fact that the
 power supplied to an outer lobe must balance the
growth rate of total energy in an outer lobe, $U_{OL}$
(including kinetic energy in the channel and 
work done on IGM).  Collecting all terms gives
\be
{ d U_{OL} \over dt}
= \pi a^2 f_V v_{end} \left( {\Gamma_c p \over \Gamma_c -1} \right)
+  {1 \over 2 }   \pi r_c^2 f_V v_{end} \rho_c v_c^2 
\left( 1 + 2 { A_{end} \over A_c} \right) 
\label{Full_energy_equation}
\ee
The first term represents the enthalpy budget of the expanding lobe, and
assumes the adiabatic index is the same for the channel and the cocoon. The 
second term describes the kinetic energy of, and work done by, the
channel flow. I have assumed the channel expands proportionally to the
full lobe ($r_c/a$ remains constant) and that the channel flow does work
over an area $A_{end}$. 

To proceed, equating $P_c = d U_{OL}/dt$, and 
using \eref{Mach_number_relation} to write the $\rho_c v_c^2$ 
term in \eref{Full_energy_equation} in terms of known quantities,
gives
\be
P_c = \pi a^2  f_V v_{end} { \Gamma_c p \over \Gamma_c -1} 
\left( 1 + {\cal C}\right)
\label{General_channel_power}
\ee
Here, the first term  is the basic power  needed to
grow the lobe against the IGM pressure, and the additional term
\be
{\cal C}
 = { r_c^2 \over a^2} { \Gamma_c -1 \over 2 \Gamma_c} 
\left( 1 + { 2 A_{end} \over A_c} \right) \left[ \Gamma_{IGM} {\mach}_{end}^2
{ A_{end} \over A_c} + \left( { A_{end} \over A_c} -1\right) \right]
\ee
is the extra power needed to maintain the channel. 
When $A_{end} \to A_c$, the expression for $P_c$ in
 \eref{General_channel_power}  reaches its lowest value.  Thus, in
general we have 
\be
P_c \gtw \pi a^2  f_V v_{end} { \Gamma_c p \over \Gamma_c -1} 
\left[ 1 +  
{ 3 r_c^2 \over a^2} { \Gamma_c -1 \over 2 \Gamma_c} \Gamma_{IGM} 
{\mach}_{end}^2 \right]
\label{Power_result}
\ee
Because $p$ and $a$ are known from the data, \eref{Power_result}
shows that  the minimum power $P_c$ required
for a specific choice of $v_{end}$ (and therefore $\mach_{end}$) depends only
on the fractional channel radius, $r_c/a$ and the volume growth factor
$f_V$.  Some numerical examples are
discussed in the text and given in Table \ref{Flow_driven_models}.

\end{appendix}

\section*{References}

\begin{harvard}

\item[]
Abdo A A \etal  (Fermi-LAT Collaboration) 2010 {\em Science} {\bf 328} 725 (A10)  

\item[]
Alvarez H, Aparici J, May J and Reich P, 2000 \AandA {\bf 355} 863

\item[]
Benford G and Protheroe R, 2008 \MNRAS {\bf 383} 663

\item[]
Bicknell G 1984 \ApJ {\bf 286} 68

\item[]
Bicknell G and Melrose, D 1982 \ApJ {\bf 262} 511

\item[]
Borovsky J and Eilek J 1986 \ApJ {\bf 308} 929

\item[]
Bouchard A, Jerjen H, Da Costa G S and Ott J 2007 {\em Astron.
J.} {\bf 133} 261

\item[]
Bosch-Ramin V 2012 \AandA {\bf 542} 125

\item[]
Brunetti G, Blasi P, Cassano R. and Gabici S 2004 \MNRAS {\bf 350} 1174

\item[]
Burbidge G 1956 \ApJ {\bf 124} 416

\item[]
Burn B J 1966, \MNRAS {\b 133} 67

\item[]
Casse F, Lemoine M and Pelletier G 2002 {\em Phys. Rev. D} {\bf 65}
 023002

\item[] Combi J and Romero G 1997 \AandA {\em Supp.} {\bf 121} 11

\item[]
Croston J H, Kraft R P, Hardcastle M J \etal  2009 \MNRAS {\bf 395}
 1999

\item[]
de Gouveia dal Pino E M and Lazarian A 2005 \AandA {\bf 441} 845

\item[]
De Young D S  1986 \ApJ {\bf 307} 62

\item[]
Doe S M, Ledlow M J, Burns J O and White R A 1995 \AJ {\bf 110} 46

\item[]
Dosch A, Shalcky A and Tautz R C 2011 \MNRAS {\bf 413} 2950

\item[]
Dreher J W, Carilli C L and Perley R A 1987 \ApJ {\bf 316} 611

\item[]
Drury L O'C 1983 {\em Reports in Progress in Physics} {\bf 46} 973

\item[]
Drury L O'C 2012 \MNRAS {\bf 422} 2474  

\item[]
Eilek J 1979 \ApJ {\bf 230} 373

\item[]
Eilek J and Arendt P  1996 \ApJ {\bf 457} 150

\item[]
Feain I J, Ekers R D, Murphy T \etal 2009 \ApJ {\bf 707} 114

\item[]
Feain I J, Cornwell T, Ekers R \etal 2011 \ApJ {\bf 740} 17

\item[]
Finoguenov A, Davis D S, Zimer M and Mulchaey J S 2006 \ApJ
{\bf 646} 143

\item[]
Garrington S T and Conway R G 1991 \MNRAS {\bf 250} 198

\item[]
Gekelman W, Lawrence E and Van Compernolle B 2012 \ApJ {\bf 753} 131

\item[]
Giacalone J 2005 \ApJ  {\bf 624} 765

\item[]
Gourgouliatos K N and Lyutikov M 2012 \MNRAS {\bf 420} 505

\item[]
Guidetti D, Laing R A, Bridle A H, \etal  2011
\MNRAS {\bf 413} 2525

\item[]
Gunn J  E and Gott J R III 1972 \ApJ {\bf 176} 1

\item[]
Hardcastle M, Cheung C, Feain I and Stawarz L 2009 \MNRAS {\bf 393}
 1041 (H09)

\item[]
Harris G L H, Rejkuba M and Harris W E 2010 {\em Pub. Astron. Soc. 
Pacific} {\bf 27} 457

\item[]
Haugen N E, Brandenburg A and Dobler W 2004 {\em Phys. Rev. E} {\bf 70} 016308

\item[]
Junkes N, Haynes F, Jarnett J and Jauncey D 1993 \AandA {\bf 269} 29

\item[]
Kagan D, Milosavljevic M and Spitkovsky A 2013 \ApJ {\bf 774} 41

\item[]
Kang H 2011 {\em J. Kor. Astron. Soc.} {\bf 44} 49

\item[]
Katz-Stone D M and  Rudnick L 1996 \ApJ {\bf 448} 146

\item[]
Katz-Stone D M, Rudnick L, Butenhogg C and O'Donoghue A A 1999
\ApJ {\bf 516} 716  

\item[]
Kowal G, de Gouveia Dal Pino E and Lazarian A 2012 {\em Phys. Rev. Letters}
{\bf 108} 241102

\item[]
Krachentsev  I D, Tully, R B, Dolphin A, \etal 2007 {\em Astron.
J.} {\bf 133} 504

\item[]
Kraft R P, Forman W R, Hardcastle M J, \etal 2009 \ApJ {\bf 698} 2036

\item[]
Lacombe, C 1977 \AandA {\bf 54} 1

\item[]
Laing R A and Bridle A H 1987 \MNRAS {\bf 228} 557

\item[]
Li H, Lovelace R, Finn J and Colgate S 2001 \ApJ {\bf 561} 915

\item[]
Litvinenko Y E 1999 \AandA {\bf 685} 690

\item[]
Lynden-Bell D 1996 \MNRAS {\bf 279} 389

\item[] 
Markidis S, Henri P, Lapenta G \etal 2012 {\em Nonlin.
Processes Geophys.} {\bf 18} 145

\item[]
McKinley B, Briggs F, Gaensler B \etal 2013 \MNRAS in press

\item[]
Mizuta A,  Yamada S and  Takabe H  2004, \ApJ {\bf 606} 804

\item[]
Molina F Z, Glover S C O, Federrath C and Klessen R S 2012 
\MNRAS {\bf 423} 2680

\item[]
Morganti R, Killeen N E B, Ekers R D and Osterloo T A 1999
\MNRAS {\bf 307} 750

\item[]
Myers S T and Spangler S R 1985 \ApJ {\bf 291} 52

\item[]
Nakamura M, Li H and Li S 2006 \ApJ {\bf 652} 2006

\item[]
Neff S, Eilek J and Owen F 2014 in preparation

\item[]
Norman M L,  Winkler K-H and Smarr, L, 1983,
in {\em Astrophysical Jets}  (eds A Ferrari and A G Pacholczyk;  Dordrecht:  
Reidel) 227; 

\item[]
O'Sullivan S, Reville B and Taylor A M 2009 \MNRAS {\bf 400} 2480

\item[]
O'Sullivan S P, Feain I J, McClure-Griffiths N M, \etal 2013
\ApJ {\bf 764} 162

\item[]
Pacholczyk A G. 1970 {\em Radio Astrophysics} (San Francisco:  Freeman)

\item[]
Quillen A C, Graham J R and Frogel J A 1993 \ApJ {\bf 412} 550

\item[]
Rejkuba M, Harris W E, Greggio L and Harris G L H 2011 \AandA {\bf 526} 123

\item[]
Retin\`o A, Sundkvist D, Viavads A, \etal 2011 {\em Nature Physics} {\bf 3}, 236

\item[]
Riquelme M A  and Spitkovsky A 2010 \ApJ {\bf 717} 1054

\item[]
Romanova M M and Lovelace R V E 1992 \AandA {\bf 262} 26

\item[]
Rudnick L and Blundell K M 2003 \ApJ {\bf 588} 143

\item[]
Schekochihin A A, Cowley S C, Taylor S F, \etal 2004 \ApJ {\bf 612} 276

\item[]
Schure K M, Bell A R, Drury L O'C and Bykov A M 2012 {\em Space. Sci. Rev.} 
{\bf 173} 491

\item[]
Servidio S, Dmitruk P, Crego A, \etal 2011 {\em Nonlin.
Processes Geophys.} {\bf 18} 675

\item[]
Shay M A, Drake J F, Swisdak M and Rogers B N 2005 {\em Phys. Plasmas.} 
{\bf 11} 2199

\item[]
Spangler, S R and Sakurai T 1985 \ApJ {\bf 297} 84

\item[]
Sparke L S 1996 \ApJ {\bf 473} 810

\item[]
Stawarz L and Petrosian V 2008 \ApJ {\bf 681} 1725

\item[]
Stawarz L, Tanaka Y T, Madejsk, G \etal 2013 \ApJ {\bf 766} 48

\item[]
Stefan I I, Carilli C L, Green D A, \etal 2013 \MNRAS in press.

\item[]
Sun M, Voit G M, Donahue, M, \etal 
2009, \ApJ {\bf 693} 1142

\item[]
Tang X Z 2008 \ApJ {\bf 679} 1000

\item[]
Tang X  Z and Boozer A H 2004 {\em Phys Plasmas} {\bf 11} 2679

\item[]
Tregillis I L, Jones T W and Ryu D 2001 \ApJ {\bf 557} 475

\item[]
Uzdensky D A and MacFadyen A I 2006, \ApJ {\bf 647} 1192

\item[]
Wilson A S, Smith D A and Young A J 2006 \ApJ {\bf 644} L9

\item[]
  Wise M W, McNamara B R, Nulsen P E J \etal  2007 \ApJ {\bf 659} 1153

\item[]
Woodley K A, Harris W E, Puzia T H \etal 2010 \ApJ {\bf 708} 1335

\item[]
Wykes S, Croston J, Hardcastle M \etal 2013 \AandA {\bf 558} A19

\item[]
Yamada M, Ji H, Hsu S \etal 1997 {\em Phys. Plasmas} {\bf 4} 1936

\item[]
Yang R-Z, Sahakyan N, de Ona Wilhelmi E, \etal
 2012 \AandA {\bf 542} A19 (Y12)

\item[]
Zhdankin V, Uzensky D A, Perez J C and Boldyrev S, 2013 \ApJ {\bf 771} 124

\item[]
Zhou Y, Matthaeus W and Dmitruk P, 2004 {\em Rev. Mod. Phys} {\bf 76} 1015

\item[]
Zweibel E G and Yamada M  2009 {\em Ann. Rev. Astron. Astophys.} {\bf 47} 129

\end{harvard}

\end{document}